\documentclass[aps,prd,amsmath,twocolumn,amssymbaps,showpacs]{revtex4-2}
\usepackage{graphicx}  
\usepackage{dcolumn}   
\usepackage{bm}        
\usepackage{amssymb}   
\usepackage{amsmath}   
\usepackage{bbm}
\usepackage{draftcopy}
\usepackage{relsize} 
\usepackage{xfrac}    
\usepackage{slashed}
\usepackage{color}
\usepackage{footnote}
\definecolor{dgreen}{cmyk}{1.,0.,1.,0.2}        
\definecolor{orange}{cmyk}{0.,0.353,1.,0.}    

\usepackage[bookmarks]{hyperref}

\newcommand{\di}{{\rm d}}

\newcommand{\tr}{{\rm tr}}

\newcommand{\be}{\begin{equation}}
\newcommand{\ee}{\end{equation}}                                                                               
\newcommand{\bea}{\begin{eqnarray}}
\newcommand{\eea}{\end{eqnarray}}

\begin{document}
\title{Kaon superfluidity in the early universe}

\author{Gaoqing Cao}
\affiliation{School of Physics and Astronomy, Sun Yat-sen University, Zhuhai 519088, China}

\date{\today}

\begin{abstract}
Previously, it was found that pion superfluidity could be realized in the QCD epoch of the early universe, when lepton flavor asymmetry $|l_{\rm e}+l_\mu|$ is large enough to generate a charge chemical potential $|\mu_{\rm Q}|$ larger than vacuum pion mass. By following the same logic, kaon superfluidity might also be possible when $|l_{\rm e}+l_\mu|$ is so large that $|\mu_{\rm Q}|$ becomes larger than vacuum kaon mass. Such a possibility is checked by adopting Ginzburg-Landau approximation within the three-flavor Polyakov--Nambu--Jona-Lasinio model. Consider the case with full chemical balance, though kaon superfluidity could be stable compared to the chiral phases with only $\sigma$ condensations, it would get killed by the more favored homogeneous pion superfluidity. If we introduce mismatch between $s$ and $d$ quarks, kaon superfluidity would require so large $s$ quark density that such a state is impossible in the early universe.
\end{abstract}


\maketitle

\section{Introduction}
One important mission of nuclear physics is to explore possible phases of quantum chromodynamics (QCD) systems under different circumstances. Usually, neutron stars are believed to be in the low energy regime of QCD with hadrons the basic degrees of freedom, and many relevant phases had been proposed for them, such as neutron Cooper pairing~\cite{Elgaroy:1996mg}, pasta structure~\cite{Ravenhall:1983uh,Hashimoto:1984pap}, pion condensation~\cite{Migdal1971,Sawyer:1972cq,Scalapino:1972fu}, and Kaon condensation~\cite{Kaplan:1986yq}, see also the review~\cite{Heiselberg:1999mq}. On the other hand, relativistic heavy ion collisions (HICs) are expected to be in the high energy regime of QCD with quarks and gluons the basic degrees of freedom, and the transition between quark gluon plasma (QGP) and hadron phases were widely studied~\cite{Yagi:2005yb,Luo:2017faz}. However, the line between QGP and hadron phases is never very clear due to two facts: no sign of ordered phase transition was ever found in HICs~\cite{Aoki:2006we,Bhattacharya:2014ara,Floris:2014pta,Adamczyk:2017iwn} and quarkyonic matter was supposed to be possible in neutron stars ~\cite{Fukushima:2015bda,McLerran:2018hbz,Cao:2020byn,Cao:2022inx,Xia:2023omv}.

The QCD epoch of the early universe is another playground of high energy nuclear physics. Previously, large lepton flavor asymmetry~\cite{Vovchenko:2020crk,Middeldorf-Wygas:2020glx} and primordial magnetic field~\cite{Vachaspati:1991nm,Son:1998my,Grasso:2000wj} were assumed to be generated in the electroweak epoch, and the phase transition in the QCD epoch could be affected strongly~\cite{Vovchenko:2020crk,Middeldorf-Wygas:2020glx,Cao:2021gfk,Cao:2022fow}. Mainly, pion superfluidity could be favored for large enough lepton flavor asymmetry~\cite{Vovchenko:2020crk,Middeldorf-Wygas:2020glx,Cao:2021gfk} and the QCD transition becomes of first order after taking the magnetic effect into account~\cite{Cao:2022fow}. The first-order transition could generate gravitational wave directly in the QCD epoch, and the European Pulsar Timing Array group considered it to be a possible origin of their updated observations~\cite{EPTA:2023xxk}. Inspired from the studies on neutron stars~\cite{Migdal1971,Sawyer:1972cq,Scalapino:1972fu,Kaplan:1986yq} and pure isospin matter~\cite{Son:2000xc,He:2005nk,He:2006tn,Cao:2016ats}, if the lepton flavor asymmetry is so large that the chemical potential of $K^\pm$ is larger than their mass, kaon condensation can possibly be realized in a similar way as pions in the early universe~\cite{Vovchenko:2020crk,Middeldorf-Wygas:2020glx,Cao:2021gfk,Cao:2022fow}. But it is also possible that pion superfluidity would kill kaon superfluidity at large charge density, in a similar way to the killing of rho superconductivity at finite isospin density~\cite{Brauner:2016lkh}. Indeed, in the case of full chemical balance, no sign of kaon superfluidity was found for however large isospin chemical potential in the three-flavor chiral perturbation theory with mesons the basic degrees of freedom~\cite{Andersen:2022zfp}.

 This work is devoted to exploring the possibility of kaon superfluidity in the early universe by adopting the three-flavor Polyakov--Nambu--Jona-Lasinio (PNJL) model with quarks and gluons the basic degrees of freedom. If, due to its internal structure, kaon mass is greatly reduced to be comparable to pion mass in the medium, as that happens in neutron stars~\cite{Kaplan:1986yq},  kaon superfluidity could probably coexist with pion superfluidity. The paper is organized as follows. In Sec.\ref{phase}, formalisms are developed for chiral phases in Sec.\ref{CP} and pion/kaon superfluidity in Sec.\ref{pion superfluidity}, respectively, for the general case with a primordial magnetic field. Specifically, the part for kaon condensation is developed by applying Ginzburg-Landau approximation based on the background of pion superfluidity. Within the subsections, the nontrivial sectors of strong interaction are presented in detail in Sec.\ref{PNJLB} and Sec.\ref{PNJL}, and the trivial sectors of electroweak interaction are simply summarized in Sec.\ref{EWB} and Sec.\ref{EW}. In Sec.\ref{numerical}, numerical results are demonstrated for the case without primordial magnetic field. Finally, a brief summary is given in Sec.\ref{summary}.
\section{The three-flavor PNJL model}\label{phase}
In this section, we adopt the three-flavor PNJL model~\cite{Fukushima:2017csk,Klevansky:1992qe,Hatsuda:1994pi} for the QCD sector and develop the formalism for a finite primordial magnetic field.
\subsection{Chiral phases in the magnetic field}\label{CP}
Usually, chiral symmetry breaking and restoration are related to the expectation values of scalar field condensates. Below, we specifically refer to such phases as {\it chiral phases} to distinguish from pion or kaon superfluidity phase by following the previous convention~\cite{Cao:2022fow}. Due to the Meissner effect, the chiral phases and pion/kaon superfluidity should be treated separately in an external magnetic field $H$, and Gibbs free energy must be adopted to determine the ground state~\cite{Fetter2003b}. As magnetization is small in chiral phases~\cite{Cao:2022fow}, we would directly use $H$ instead of magnetic induction intensity $B$ in the following.
\subsubsection{The strong interaction sector}\label{PNJLB}
In a primordial magnetic field, the Lagrangian of the PNJL model can be modified from the previous one~\cite{Cao:2022fow} to
\begin{eqnarray}
{\cal L}_{\rm PNJL}\!&=&-{H^2\over 2}\!+\!\bar\psi\!\left[i\slashed{D}\!-\!i\gamma^4\!\!\left(\!ig{\cal A}^4\!+\!Q_{\rm q}\mu_{\rm Q}\!+\!{\mu_{\rm B}\over3}\!\right)\!-\!m_0\right]\!\psi\nonumber\\
&&\!\!\!+G\sum_{a=0}^8\!\left[(\bar\psi\lambda^a\psi)^2\!+\!(\bar\psi i\gamma_5\lambda^a\psi)^2\right]\!+\!{\cal L}_{\rm tH}\!-\!V(L)\label{LH}
\end{eqnarray}
by adopting the covariant derivative $D_\mu=\partial_\mu+i\,Q_{\rm q}eA_\mu$. Here, the field variables are defined as the following: $\psi=(u,d,s)^T$ is the three-flavor quark field, $H$ is the magnetic field with $A_\mu$ the corresponding vector potential, and the Polyakov loop is $L={1\over N_{\rm c}}\tr\,e^{ig\int\di x_4{\cal A}^4}$ with ${\cal A}^4=A^{\rm 4c}T^{\rm c}/2$ the non-Abelian gauge field. For the quarks, the current mass and electric charge number matrices are, respectively,
 \bea
 m_0&\equiv&{\rm diag}(m_{\rm 0u},m_{\rm 0d},m_{\rm 0s}),\nonumber\\
 Q_{\rm q}&\equiv&{\rm diag}(q_{\rm u},q_{\rm d},q_{\rm s})={1\over3}{\rm diag}(2,-1,-1);
 \eea
and the interaction vertices $\lambda^i~(i=1,\dots,8)$ are Gell-Mann matrices in flavor space and $\lambda^0=\sqrt{2/3}~\mathbbm{1}_3$. For later use, the 't Hooft term, ${\cal L}_{\rm tH}\equiv-K\sum_{t=\pm}{\rm Det}~\bar\psi\Gamma^t\psi$, can be rewritten as
\bea
\!\!\!\!\!\!\!\!{\cal L}_{\rm tH}\!=\!-{K\over2}\sum_{t=\pm}\epsilon_{ijk}\epsilon_{imn}(\bar{\psi}^i\Gamma^t{\psi}^i)(\bar{\psi}^j\Gamma^t{\psi}^m)(\bar{\psi}^k\Gamma^t{\psi}^n)
\eea
with $\Gamma^\pm=\mathbbm{1}_4\pm\gamma_5$ for right- and left-handed channels, respectively. Here, one should note the correspondences between $1,2,3$ and $u,d,s$ and the Einstein summation convention for the flavor indices $i,j,k,m,n$. The pure gluon potential could be obtained by fitting to the lattice QCD data~\cite{Fukushima:2017csk}:
 \bea
{V(L)\over T^4}&=&-{1\over2}\left(3.51-{2.47\over\tilde{T}}+{15.2\over\tilde{T}^{2}}\right)L^2-{1.75\over\tilde{T}^{3}}\nonumber\\
&&\times\ln\left[1-6\,L^2+8\,L^3-3L^4\right],
\eea
where $\tilde{T}\equiv T/T_0$ is the reduced temperature with $T_0=0.27\,{\rm GeV}$.

In the chiral phases, we only consider chiral condensates $\sigma_{\rm i}\equiv\langle\bar{\psi}^i{\psi}^i\rangle$ with $i$ flavor index, then the 't Hooft term ${\cal L}_{\rm tH}$ can be effectively reduced to four fermion interaction terms in Hartree approximation~\cite{Klevansky:1992qe}:
\begin{widetext}
\bea
{\cal L}_{\rm tH}^4&=&-{K\over2}\sum_{s=\pm}\epsilon_{ijk}\epsilon_{imn}\langle\bar{\psi}^i\Gamma^s{\psi}^i\rangle(\bar{\psi}^j\Gamma^s{\psi}^m)(\bar{\psi}^k\Gamma^s{\psi}^n)\nonumber\\
&=&-{K\over6}\Big\{2\sum_{\rm f=u,d,s}\sigma_{\rm f}(\bar{\psi}\lambda^0\psi)^2-3\sigma_s\sum_{i=1}^3(\bar{\psi}\lambda^i\psi)^2
-3\sigma_{\rm d}\sum_{i=4}^5(\bar{\psi}\lambda^i\psi)^2-3\sigma_{\rm u}\sum_{i=6}^7(\bar{\psi}\lambda^i\psi)^2+(\sigma_s\!-\!2\sigma_{\rm u}\!-\!2\sigma_{\rm d})(\bar{\psi}\lambda^8\psi)^2\nonumber\\
&&+\sqrt{2}(2\sigma_s\!-\!\sigma_{\rm u}\!-\!\sigma_{\rm d})(\bar{\psi}\lambda^0\psi)(\bar{\psi}\lambda^8\psi)-\sqrt{6}(\sigma_{\rm u}\!-\!\sigma_{\rm d})(\bar{\psi}\lambda^3\psi)(\bar{\psi}\lambda^0\psi-\sqrt{2}\bar{\psi}\lambda^8\psi)\Big\}-(\lambda^a\rightarrow i\lambda^a\gamma^5)
\eea
with $\epsilon_{ijk}$ the Levi-Civita symbol. As a consequence, the Lagrangian becomes the one with only four fermion effective interactions,
\begin{eqnarray}
\!\!\!\!\!\!{\cal L}_{\rm PNJL}^4\!\!=\!-{H^2\over 2}\!-\!V(L)\!+\!\bar\psi\!\left[i\slashed{D}\!-\!i\gamma^4\left(\!ig{\cal A}^4\!+\!Q_{\rm q}\mu_{\rm Q}\!+\!{\mu_{\rm B}\over3}\!\right)\!-\!m_0\right]\!\psi\!+\!\!\!\sum_{a,b=0}^8\!\!\left[G_{ab}^-(\bar\psi\lambda^a\psi)(\bar\psi\lambda^b\psi)\!+\!G_{ab}^+(\bar\psi i\gamma_5\lambda^a\psi)(\bar\psi i\gamma_5\lambda^b\psi)\right]\!,
\end{eqnarray}
where the nonvanishing elements of the symmetric coupling matrices $G^\pm$ are given by~\cite{Klevansky:1992qe}
\begin{eqnarray}
&&G_{00}^\mp=G\mp {K\over3}\sum_{\rm f=u,d,s}\sigma_{\rm f},~G_{11}^\mp=G_{22}^\mp=G_{33}^\mp=G\pm {K\over2}\sigma_s,~G_{44}^\mp=G_{55}^\mp=G\pm {K\over2}\sigma_{\rm d},~G_{66}^\mp=G_{77}^\mp=G\pm {K\over2}\sigma_{\rm u},\nonumber\\
&&G_{88}^\mp=G\mp {K\over6}(\sigma_s-2\sigma_{\rm u}-2\sigma_{\rm d}),~G_{08}^\mp=\mp {\sqrt{2}K\over12}(2\sigma_s\!-\!\sigma_{\rm u}\!-\!\sigma_{\rm d}),~G_{38}^\mp=-\sqrt{2}G_{03}^\mp=\mp {\sqrt{3}K\over6}(\sigma_{\rm u}\!-\!\sigma_{\rm d}).
\end{eqnarray}

By contracting a pair of field and conjugate field further in ${\cal L}_{\rm tH}^4$, we find
\begin{eqnarray}
{\cal L}_{\rm tH}^2&=&-\sum_{s=\pm}^{i(\neq j\neq k)}K\langle\bar{\psi}^j\Gamma^s{\psi}^j\rangle\langle\bar{\psi}^k\Gamma^s{\psi}^k\rangle[\bar{\psi}^i\Gamma^s{\psi}^i]=-2K\sigma_j\sigma_k\bar{\psi}^i{\psi}^i\  (i\neq j\neq k, j<k),
\end{eqnarray}
and we could obtain the total dynamical quark masses as
\begin{eqnarray}\label{massi}
m_i&=&m_{0i}-4G\sigma_i+2K\sigma_j\sigma_k
\end{eqnarray}
after taking the initial four quark interactions into account. Then, the gap equations directly follow the definitions of chiral condensates, that is,
\begin{eqnarray}
\sigma_i\equiv\langle\bar{\psi}^i{\psi}^i\rangle=-{i\over V_4}{\rm Tr}~{\cal S}_{i},
\end{eqnarray}
where the effective quark propagators are given by~\cite{Cao:2021rwx}
\begin{eqnarray}
\hat{{\cal S}}_{\rm i}({k})
&=&i\int {\di s}\exp\Big\{-i (m_{\rm i}^{2}+{k}_4^2+k_3^2)s-i{\tan(q_{\rm i}eHs)\over q_{\rm i}eH}(k_1^2+k_2^2)\Big\}\left[m_{\rm i}-\gamma^4k_4\!-\!\gamma^3k_3\!-\!\gamma^2(k_2+{\tan(q_{\rm i}eHs)}k_1)\right.\nonumber\\
&&\left.-\gamma^1(k_1-{\tan(q_{\rm i}eHs)}k_2)\right]\Big[1
+{\gamma^1\gamma^2\tan(q_{\rm i}eHs)}\Big]\label{prop_m}
\end{eqnarray}
for the case with a constant magnetic field.

By adopting vacuum regularization, the explicit forms of the gap equations are~\cite{Cao:2021rwx}
\begin{eqnarray}
-\sigma_{\rm f}
&=&N_c{m_{\rm f}^3\over2\pi^2}\Big[\tilde{\Lambda}_{\rm f}\Big({1+\tilde{\Lambda}_{\rm f}^2}\Big)^{1\over2}-\ln\Big({\tilde\Lambda_{\rm f}}
+\Big({1+\tilde{\Lambda}_{\rm f}^2}\Big)^{1\over2}\Big)\Big]+N_c{m_{\rm f}\over4\pi^2}\int_0^\infty {ds\over s^2}e^{-m_{\rm f}^2s}\left({q_{\rm f}eHs
	\over\tanh(q_{\rm f}eHs)}-1\right)\nonumber\\
&&-6\sum_{\rm u=\pm}{|q_{\rm f}eH|\over 2\pi}\sum_{n=0}^\infty\alpha_{\rm n}\int_{-\infty}^\infty {\di  k_3\over2\pi}{m_{\rm f}
\over E_{\rm f}^{\rm n}}F_{\rm f}^{\rm u}(E_{\rm f}^{\rm n},L,T,\mu_{\rm Q},\mu_{\rm B}),\label{mgap}
\end{eqnarray}
where the reduced cutoff $\tilde{\Lambda}_{\rm f}={\Lambda/m_{\rm f}}$, Landau level factor $\alpha_{\rm n}=1-\delta_{\rm n0}/2$, particle energy $E_{\rm f}^{\rm n}(k_3,m_{\rm f})=(2n|q_{\rm f}eH|+k_3^2+m_{\rm f}^2)^{1/2}$, and the fermion distribution function $$F_{\rm f}^{\rm u}(E_{\rm f}^{\rm n},L,T,\mu_{\rm Q},\mu_{\rm B})\equiv{L\,e^{-{1\over T}\left(E_{\rm f}^{\rm n}-u\left(q_{\rm f}\mu_{\rm Q}+{\mu_{\rm B}\over 3}\right)\right)}+2L\,e^{-{2\over T}\left(E_{\rm f}^{\rm n}-u\left(q_{\rm f}\mu_{\rm Q}+{\mu_{\rm B}\over 3}\right)\right)}+e^{-{3\over T}\left(E_{\rm f}^{\rm n}-u\left(q_{\rm f}\mu_{\rm Q}+{\mu_{\rm B}\over 3}\right)\right)}\over 1+3L\,e^{-{1\over T}\left(E_{\rm f}^{\rm n}-u\left(q_{\rm f}\mu_{\rm Q}+{\mu_{\rm B}\over 3}\right)\right)}+3L\,e^{-{2\over T}\left(E_{\rm f}^{\rm n}-u\left(q_{\rm f}\mu_{\rm Q}+{\mu_{\rm B}\over 3}\right)\right)}+e^{-{3\over T}\left(E_{\rm f}^{\rm n}-u\left(q_{\rm f}\mu_{\rm Q}+{\mu_{\rm B}\over 3}\right)\right)}}.$$ Eventually, the quark part of thermodynamic potential can be consistently obtained as~\cite{Cao:2023bmk}
	\begin{eqnarray}
	\Omega_{\rm q}(H)&=&2G\sum_{{\rm f}=u,d,s}\sigma_{\rm f}^2-4K\prod_{{\rm f}=u,d,s}\sigma_{\rm f}-N_c\sum_{{\rm f}=u,d,s}\left\{{m_{\rm f}^4\over8\pi^2}\Big[\tilde{\Lambda}_{\rm f}\Big(1+{2\tilde{\Lambda}_{\rm f}^2}\Big)\Big({1+{\tilde{\Lambda}_{\rm f}^2}}\Big)^{1\over2}-\ln\Big({\tilde{\Lambda}_{\rm f}}
+\Big({1+{\tilde{\Lambda}_{\rm f}^2}}\Big)^{1\over2}\Big)\Big]\right.\nonumber\\
&&-{1\over8\pi^2}\int_0^\infty {ds\over s^3}\left(e^{-m_{\rm f}^2s}-e^{-{m_{\rm f}^{\rm v}}^2s}\right)\left({q_{\rm f}eHs
	\over\tanh(q_{\rm f}eHs)}-1\right)-{1\over8\pi^2}\int_0^\infty {ds\over s^3}e^{-{m_{\rm f}^{\rm v}}^2s}\left({q_{\rm f}eHs
	\over\tanh(q_{\rm f}eHs)}-1-{1\over 3}(q_{\rm f}eHs)^2\right)\nonumber\\
&&\left.+2T\sum_{\rm u=\pm}{|q_{\rm f}eH|\over 2\pi}\sum_{n=0}^\infty\alpha_{\rm n}\int_{-\infty}^\infty {\di  k_3\over2\pi}K_{\rm f}^{\rm u}(E_{\rm f}^{\rm n},L,T,\mu_{\rm Q},\mu_{\rm B})\right\},\label{Omgq}
\end{eqnarray}
with
$$K_{\rm f}^{\rm u}(E_{\rm f}^{\rm n},L,T,\mu_{\rm Q},\mu_{\rm B})={1\over N_{\rm c}}\ln\left[1+3L\,e^{-{1\over T}\left(E_{\rm f}^{\rm n}-t\left(q_{\rm f}\mu_{\rm Q}+{\mu_{\rm B}\over 3}\right)\right)}+3L\,e^{-{2\over T}\left(E_{\rm f}^{\rm n}-t\left(q_{\rm f}\mu_{\rm Q}+{\mu_{\rm B}\over 3}\right)\right)}+e^{-{3\over T}\left(E_{\rm f}^{\rm n}-t\left(q_{\rm f}\mu_{\rm Q}+{\mu_{\rm B}\over 3}\right)\right)}\right].$$
And the Gibbs free energy for the QCD sector is $\Omega_{\rm \chi}^{\rm M}=-{H^2\over 2}+V(L)+\Omega_{\rm q}(H)$~\cite{Cao:2022fow}.

With the free energy at hand, the gap equation for $L$ can be given by $\partial_{\rm L}\Omega_{\rm \chi}^{\rm M}=0$ as
\bea
&&T^3\!\!\left[-\left(3.51\!-{2.47\over\tilde{T}}\!+\!{15.2\over\tilde{T}^{2}}\right)L\!+\!{1.75\over\tilde{T}^{3}}{12L(1-L)^2\over1\!-\!6L^2\!+\!8L^3\!-\!3L^4}\right]={6}\sum_{\rm f=u,d,s}\sum_{\rm u=\pm}{|q_{\rm f}eH|\over 2\pi}\sum_{n=0}^\infty\alpha_{\rm n}\int_{-\infty}^\infty {\di  k_3\over2\pi}\nonumber\\
&&{\,e^{-{1\over T}\left(E_{\rm f}^{\rm n}-u\left(q_{\rm f}\mu_{\rm Q}+{\mu_{\rm B}\over 3}\right)\right)}+\,e^{-{2\over T}\left(E_{\rm f}^{\rm n}-u\left(q_{\rm f}\mu_{\rm Q}+{\mu_{\rm B}\over 3}\right)\right)}\over 1+3L\,e^{-{1\over T}\left(E_{\rm f}^{\rm n}-u\left(q_{\rm f}\mu_{\rm Q}+{\mu_{\rm B}\over 3}\right)\right)}+3L\,e^{-{2\over T}\left(E_{\rm f}^{\rm n}-u\left(q_{\rm f}\mu_{\rm Q}+{\mu_{\rm B}\over 3}\right)\right)}+e^{-{3\over T}\left(E_{\rm f}^{\rm n}-u\left(q_{\rm f}\mu_{\rm Q}+{\mu_{\rm B}\over 3}\right)\right)}},
\eea
and the entropy, electric charge, and baryon densities follow the well-known thermodynamic relations as
\bea
\!\!\!\!\!\!s_{\rm \chi}^{\rm M}&=&2N_{\rm c}\sum_{\rm f=u,d,s}\sum_{\rm u=\pm}{|q_{\rm f}eH|\over 2\pi}\sum_{n=0}^\infty\alpha_{\rm n}\int_{-\infty}^\infty {\di  k_3\over2\pi}\left[K_{\rm f}^{\rm u}(E_{\rm f}^{\rm n},L,T,\mu_{\rm Q},\mu_{\rm B})+{1\over T}\left(E_{\rm f}^{\rm n}-u\left(q_{\rm f}\mu_{\rm Q}+{\mu_{\rm B}\over 3}\right)\right)F_{\rm f}^{\rm u}(E_{\rm f}^{\rm n},L,T,\mu_{\rm Q},\mu_{\rm B})\right]\nonumber\\
&&+T^3\left\{{1\over2}\left(4\times3.51-3\times{2.47\over\tilde{T}}+2\times{15.2\over\tilde{T}^{2}}\right)L^2+{1.75\over\tilde{T}^{3}}\ln\left[1-6L^2+8L^3-3L^4\right]\right\},\label{s_B}\\
\!\!\!\!\!\!n_{\rm Q}^{\rm q,M}&=&2N_{\rm c}\sum_{\rm f=u,d,s}\sum_{\rm u=\pm}{|q_{\rm f}eH|\over 2\pi}\sum_{n=0}^\infty\alpha_{\rm n}\int_{-\infty}^\infty {\di  k_3\over2\pi}u\,q_{\rm f}F_{\rm f}^{\rm u}(E_{\rm f}^{\rm n},L,T,\mu_{\rm Q},\mu_{\rm B}),\label{nQ_B}\\
\!\!\!\!\!\!n_{\rm B}^{\rm M}&=&2\sum_{\rm f=u,d,s}\sum_{\rm u=\pm}{|q_{\rm f}eH|\over 2\pi}\sum_{n=0}^\infty\alpha_{\rm n}\int_{-\infty}^\infty {\di  k_3\over2\pi}uF_{\rm f}^{\rm u}(E_{\rm f}^{\rm n},L,T,\mu_{\rm Q},\mu_{\rm B}).\label{nB_B}
\eea

\subsubsection{The electroweak interaction sector}\label{EWB}
In free gas approximation, the thermodynamic potentials for the quantum electroweak dynamics (QEWD) sector can be simply given by~\cite{Kapusta2006,Schwinger:1951nm}
\bea
\Omega_\gamma&=&2T\int{\di^3k\over(2\pi)^3}\log\left(1-e^{- k/T}\right),\\
\Omega_{\rm l}^{\rm M}&=&\sum^{\rm i=e,\mu,\tau}\left\{-T\sum_{u=\pm}\int{\di^3k\over(2\pi)^3}\log\left[1+e^{- (k-u\,\mu_{\rm i})/T}\right]+{1\over8\pi^2}\int_0^\infty{\di s\over s^3}e^{-m_{\rm i}^2s}\left[{eHs\over\tanh(eHs)}-1-{1\over3}(eHs)^2\right]\right.\nonumber\\
&&\left.-2T\sum_{u=\pm}{|eH|\over 2\pi}\sum_{n=0}^\infty\alpha_{\rm n}\int_{-\infty}^\infty {\di  k_3\over2\pi}\log\left[1+e^{- \left(\epsilon_{\rm i}^{\rm n}(k_3,eH)-u\,(-\mu_{\rm Q}+\mu_{\rm i})\right)/T}\right]\right\},\label{OmegaBl}
\eea
where $\epsilon_{\rm i}^{\rm n}(k_3,eH)=(k_3^2+2n|eH|+m_{\rm i}^2)^{1/2}$ and the degeneracy is one for (anti-)neutrinos due to their definite chiralities. Then, the corresponding entropy, electric charge and lepton flavor densities follows directly as
\bea
s_\gamma&=&2\int{\di^3k\over(2\pi)^3}\left[-\log\left(1-e^{- k/T}\right)+{k/T\over e^{k/T}-1}\right],\\
s_{\rm l}^{\rm M}&=&\sum^{\rm i=e,\mu,\tau}_{u=\pm}\left\{\int{\di^3k\over(2\pi)^3}
\left\{\log\left[1+e^{- (k-u\,\mu_{\rm i})/T}\right]+{(k-u\,\mu_{\rm i})/T\over1+e^{(k-u\,\mu_{\rm i})/T}}\right\}+2\sum_{u=\pm}{|eH|\over 2\pi}\sum_{n=0}^\infty\alpha_{\rm n}\int_{-\infty}^\infty {\di  k_3\over2\pi}\right.\nonumber\\
&&\left.\left\{\log\left[1+e^{- \left(\epsilon_{\rm i}^{\rm n}(k_3,eH)-u\,(-\mu_{\rm Q}+\mu_{\rm i})\right)/T}\right]+{\left(\epsilon_{\rm i}^{\rm n}(k_3,eH)-u\,(-\mu_{\rm Q}+\mu_{\rm i})\right)/T\over1+e^{\left(\epsilon_{\rm i}^{\rm n}(k_3,eH)-u\,(-\mu_{\rm Q}+\mu_{\rm i})\right)/T}}\right\}\right\},\\
n_{\rm Q}^{\rm l,M}&=&2T\sum_{u=\pm}{|eH|\over 2\pi}\sum_{n=0}^\infty\alpha_{\rm n}\int_{-\infty}^\infty {\di  k_3\over2\pi}{-u\over1+e^{\left(\epsilon_{\rm i}^{\rm n}(k_3,eH)-u\,(-\mu_{\rm Q}+\mu_{\rm i})\right)/T}},\\
n_{\rm i}^{\rm M}&=&T\sum_{u=\pm}\int{\di^3k\over(2\pi)^3}{u\over1+e^{(k-u\,\mu_{\rm i})/T}}+2T\sum_{u=\pm}{|eH|\over 2\pi}\sum_{n=0}^\infty\alpha_{\rm n}\int_{-\infty}^\infty {\di  k_3\over2\pi}{u\over1+e^{\left(\epsilon_{\rm i}^{\rm n}(k_3,eH)-u\,(-\mu_{\rm Q}+\mu_{\rm i})\right)/T}}.
\eea
\end{widetext}

So, in the chiral phases, the total thermodynamic potential, entropy, electric charge and lepton densities are, respectively, 
\bea
\Omega_{\rm M}&=&\Omega_\gamma+\Omega_{\rm l}^{\rm M}+\Omega_{\rm \chi}^{\rm M},\,
s_{\rm M}=s_\gamma\!+\!s_{\rm l}^{\rm M}\!+\!s_{\rm \chi}^{\rm M},\nonumber\\
n_{\rm Q}^{\rm M}&=&n_{\rm Q}^{\rm l,M}\!+\!n_{\rm Q}^{\rm q,M},\ \ \ \ \ \ \ \,
n_{\rm l}^{\rm M}=\!\sum_{\rm i=e,\mu,\tau} \!\! n_{\rm i}^{\rm M}
\eea
after combining the QEWD and QCD sectors together. According to the conventions~\cite{Vovchenko:2020crk,Middeldorf-Wygas:2020glx}, the following reduced quantities are frequently utilized:
\bea
b^{\rm M}=n_{\rm B}^{\rm M}/s_{\rm M},\ l^{\rm M}=n_{\rm l}^{\rm M}/s_{\rm M},\ l_{\rm i}^{\rm M}=n_{\rm i}^{\rm M}/s_{\rm M}.
\eea

\subsection{The superconducting pion or kaon superfluidity}\label{pion superfluidity}
Since a homogeneous pion or kaon superfluid is also a type-I electric superconductor, the external magnetic field will be entirely expelled from the bulk due to the Meissner effect~\cite{Fetter2003b}. Then, the formalism is as if no external magnetic field exists, see that for the pion superfluidity in Ref.~\cite{Cao:2022fow}. In this section, we would devote more effort to deriving the thermodynamic potential for kaon condensate based on the pion superfluidity.
\subsubsection{The strong interaction sector}\label{PNJL}
 Without magnetic field in the bulk, the Lagrangian is given by~\cite{Fukushima:2017csk,Klevansky:1992qe,Hatsuda:1994pi}:
\begin{eqnarray}
{\cal L}_{\rm PNJL}\!&=&\!-\!V(L)\!+\!\bar\psi\!\left[i\slashed{\partial}\!-\!i\gamma^4\!\!\left(\!ig{\cal A}^4\!+\!Q_{\rm q}\mu_{\rm Q}\!+\!{\mu_{\rm B}\over3}\!\right)\!-\!m_0\right]\!\psi\nonumber\\
&&+G\sum_{a=0}^8\left[(\bar\psi\lambda^a\psi)^2+(\bar\psi i\gamma_5\lambda^a\psi)^2\right]+{\cal L}_{\rm tH}.
\end{eqnarray}
For the pion or kaon superfluidity, we choose the following scalar and charged pseudoscalar condensates to be nonzero: 
\bea
&&\sigma_{\rm f}=\langle\bar\psi_{\rm f}\psi_{\rm f}\rangle,\ \Delta_\pi=\langle\bar{u}i\gamma^5d\rangle,\ \Delta_\pi^*=\langle\bar{d}i\gamma^5u\rangle,\nonumber\\
&&  \Delta_K=\langle\bar{u}i\gamma^5s\rangle,\ \Delta_K^*=\langle\bar{s}i\gamma^5u\rangle.
\eea
Without loss of generality, we set $\Delta_\pi=\Delta_\pi^*$ and $\Delta_K=\Delta_K^*$ in the following. To facilitate the study, we would like first to reduce ${\cal L}_{\rm tH}$ to an effective form with four-fermion interactions at most. By applying the Hartree approximation to contract a pair of quark and antiquark in each six-fermion interaction term~\cite{Klevansky:1992qe,Cao:2022fow}, we immediately find
\begin{eqnarray}
{\cal L}_{\rm tH}^4
\!&=&\!-{K}\!\left\{\epsilon_{ijk}\epsilon_{imn}\sigma_i\!\left(\bar{\psi}^j{\psi}^m\bar{\psi}^k{\psi}^n
\!-\!\bar{\psi}^ji\gamma^5{\psi}^m\bar{\psi}^ki\gamma^5{\psi}^n\right)\right.\nonumber\\
&&\!\!\!+2\Delta_\pi\!\!\left[\bar{s}{s}\!\left(\bar{u}i\gamma^5d\!+\!\bar{d}i\gamma^5u\!-\!\Delta_\pi\right)\!+\!\bar{s}i\gamma^5{s}\left(\bar{u}d\!+\!\bar{d}u\right)\right]\nonumber\\
&&\left.\!\!\!+2\Delta_K\!\!\left[\bar{d}{d}\!\left(\bar{u}i\gamma^5s\!+\!\bar{s}i\gamma^5u\!-\!\Delta_K\right)\!+\!\bar{d}i\gamma^5{d}\left(\bar{u}s\!+\!\bar{s}u\right)\right]\right\},\nonumber\\
\end{eqnarray}
 where the second and third terms in the brace are induced by $\pi^\pm$ and $K^\pm$ condensations, respectively. Armed with the reduced Lagrangian density:
\begin{eqnarray}\label{LNJL4}
{\cal L}_{\rm PNJL}\!\!&=&\!\!-V(L,L)\!+\!\bar\psi\!\left[i\slashed{\partial}\!-\!i\gamma^4\!\!\left(\!ig{\cal A}^4\!+\!Q_{\rm q}\mu_{\rm Q}\!+\!{\mu_{\rm B}\over3}\!\right)\!-\!m_0\right]\!\psi\nonumber\\
&&+G\sum_{a=0}^8\left[(\bar\psi\lambda^a\psi)^2+(\bar\psi i\gamma_5\lambda^a\psi)^2\right]+{\cal L}_{\rm tH}^4,
\end{eqnarray}
the left calculations can just follow that of two-flavor case in principle. 

By contracting quark and antiquark pairs once more in the interaction terms of Eq.\eqref{LNJL4}, we obtain the quark bilinear as
\begin{eqnarray}\label{LNJL2}
{\cal L}_{\rm PNJL}^2&=&\bar\psi\left[i\slashed{\partial}-i\gamma^4\left(ig{\cal A}^4+Q_{\rm q}\mu_{\rm Q}+{\mu_{\rm B}\over3}\right)-m\right.\nonumber\\
&&\left.\ \ \ \ \ -i\gamma^5\left(\lambda^1\Pi +\lambda^4{\cal K}\right)\right]\psi,
\end{eqnarray}
where the scalar and pseudoscalar masses are, respectively,
\begin{eqnarray}
&&m_i=m_{0i}-4G\sigma_i+2K(\sigma_j\sigma_k+\Delta_\pi^2\delta_{i3}+\Delta_K^2\delta_{i2}),\nonumber\\
&&\Pi =(-4G+2K\sigma_3)\Delta_\pi,\ {\cal K}=(-4G+2K\sigma_2)\Delta_K
\end{eqnarray}
with $i\neq j\neq k$. Due to the non-diagonal nature of the pseudoscalar masses $\Pi$ and ${\cal K}$, the full propagator of $\psi$ cannot be further reduced and it is hard to derive the eigenenergy analytically. Since $m_\pi\ll m_K$ in the vacuum and $\mu_{\rm d}=\mu_{\rm s}$, we expect $\Delta_K\ll \Delta_\pi$ in the superfluid phase, so it is a good approximation to neglect $\Delta_K$ for $u$ and $d$ quarks and assume $m_u=m_d\equiv m_{\rm l}$. Eventually, to give consistent result on the kaon vacuum mass, we would adopt the following approximate scalar and pseudoscalar masses for further study:
\begin{eqnarray}
&&m_{\rm l}=m_{\rm 0l}-4G\sigma_{\rm l}+2K\sigma_{\rm l}\sigma_{\rm s},\nonumber\\
&&m_{\rm s}=m_{\rm 0s}-4G\sigma_{\rm s}+2K(\sigma_{\rm l}\sigma_{\rm s}+\Delta_\pi^2),\nonumber\\
&&\Pi =(-4G+2K\sigma_{\rm s})\Delta_\pi,\ {\cal K}=(-4G+2K\sigma_{\rm l})\Delta_K,\label{mDelta}
\end{eqnarray}
where $m_{\rm l}, m_{\rm s},$ and $\Pi$ are exactly the same as those without kaon condensation~\cite{Cao:2021gfk}. However, even with those simplifications, it is not possible to work out a simple analytic form for the bilinear thermodynamic potential from \eqref{LNJL2}, that is,
\bea
\Omega_{\rm bl}&=&-{1\over V_4}{\rm Tr} \ln\left[\slashed{k}-i\gamma^4\left(ig{\cal A}^4+Q_{\rm q}\mu_{\rm Q}+{\mu_{\rm B}\over3}\right)-m\right.\nonumber\\
&&\left.\ \ \ \ \ -i\gamma^5\left(\lambda^1\Pi +\lambda^4{\cal K}\right)\right]
\eea
in energy momentum space. 

Inspired from the fact that only one local minimum of thermodynamic potential is involved with respect to $\Delta_\pi$ in the pion superfluidity~\cite{Cao:2022fow}, the same fact should be true with respect to $\Delta_K$ in the kaon superfluidity. Hence, $\Delta_K$ is expected to change smoothly within pion superfluidity, and we could take Ginzburg-Landau approximation to expand $\Omega_{\rm bl}$ in orders of ${\cal K}$ and calculate to a convergent order. Then, the thermodynamic potential can be separated into two parts, $\Omega_{\rm bl}=\Omega_{\rm \pi s}'+\Omega_{\rm K}'$, where
\bea
\Omega_{\pi s}'&\equiv&-{1\over V_4}{\rm Tr} \ln S^{-1}_{\rm \pi s},\\
\Omega_{\rm K}'&\equiv&-{1\over V_4}{\rm Tr} \ln\left(1-S_{\rm \pi s}i\gamma^5\lambda^4{\cal K}\right)\nonumber\\
&=&{1\over V_4}\sum_{n=1}{1\over n}{\rm Tr} \left(S_{\rm \pi s}i\gamma^5\lambda^4{\cal K}\right)^{n}
\eea
are the ones with pion and kaon condensations, respectively. Here, $S_{\rm \pi s}$ is the three-flavor fermion propagator in the pion superfluidity with its inverse defined by~\cite{Cao:2021gfk,Cao:2022fow}
\bea
S^{-1}_{\rm \pi s}&\equiv&\slashed{k}\!-\!i\gamma^4\left(ig{\cal A}^4\!+\!Q_{\rm q}\mu_{\rm Q}\!+\!{\mu_{\rm B}\over3}\right)\!-\!m\!-\!i\gamma^5\lambda^1\Pi.
\eea
Only $u$ and $d$ quarks couple with each other in $S^{-1}_{\rm \pi s}$, so the propagator of $s$ quark can be directly given by
\bea
S_{\rm s}&=&\left[\slashed{k}\!-\!i\gamma^4\left(ig{\cal A}^4\!+\!Q_{\rm s}\mu_{\rm Q}\!+\!{\mu_{\rm B}\over3}\right)\!-\!m_s\right]^{-1}.
\eea
For the $u$ and $d$ quark parts, the matrix elements of $S_{\rm \pi s}$ had been worked out to be~\cite{He:2005nk}
\bea
S_{\rm uu}&=&\sum_{t=\pm}{\tilde{k}_0+t\,\epsilon_{\rm l}({\bf k})-{\mu_{\rm Q}\over2}\over \tilde{k}_0^2-(E_{\bf k}^{\rm -t})^2}\Lambda_{\rm t}\gamma^0,\label{Suu}
\\
S_{\rm dd}&=&\sum_{t=\pm}{\tilde{k}_0+t\,\epsilon_{\rm l}({\bf k})+{\mu_{\rm Q}\over2}\over \tilde{k}_0^2-(E_{\bf k}^{\rm t})^2}\Lambda_{\rm t}\gamma^0,\\
S_{\rm ud}&=&\sum_{t=\pm}{-i\Pi\over \tilde{k}_0^2-(E_{\bf k}^{\rm -t})^2}\Lambda_{\rm t}\gamma^5,\\
S_{\rm du}&=&\sum_{t=\pm}{-i\Pi\over \tilde{k}_0^2-(E_{\bf k}^{\rm t})^2}\Lambda_{\rm t}\gamma^5
\eea
with the help of massive energy projectors
\bea
\Lambda_\pm={1\over2}\left(1\pm\gamma^0{\boldsymbol{\gamma\cdot k}+m_{\rm l}\over \epsilon_{\rm l}({\bf k})}\right).
\eea
Here, we have defined $\tilde{k}_0\equiv k_0+i\,g{\cal A}^4+{\mu_{\rm Q}\!+\!2\mu_{\rm B}\over6}$ and $E_{\bf k}^{\rm t}\!=\!\sqrt{\left[\epsilon_{\rm l}(k)\!+t{\mu_{\rm Q}\over2}\right]^2\!+\Pi^2}$ with $\epsilon_{\rm i}(k)\!=\!\sqrt{k^2\!+m_{\rm i}^2}$. Note the orthogonality: $\Lambda_{\rm t_1}\Lambda_{\rm t_2}=\Lambda_{\rm t_1}\delta_{\rm t_1,t_2}$.

The explicit form of the thermodynamic potential had been worked out for the pion superfluidity in our previous works~\cite{Cao:2021gfk,Cao:2022fow}, refer to Eq.~\eqref{Omgpis}.
We now focus on deriving the explicit expression for the kaon part, $\Omega_{\rm K}'$. As the $s$ quark propagator is diagonal in $S_{\rm \pi s}$, it is easy to check that odd terms of ${\cal K}$ do not contribute in $\Omega_{\rm K}'$, thus we have
\bea
\Omega_{\rm K}'={1\over V_4}\sum_{n=1}{1\over n}{\rm Tr} \left(S_{\rm s}i\gamma^5S_{\rm uu}i\gamma^5\right)^{n}{\cal K}^{2n}.
\eea
If we define an effective $s$-quark propagator
\bea
S_{\rm 5s5}\equiv i\gamma^5S_{\rm s}i\gamma^5=\gamma^0\sum_{\rm u=\pm}{\Lambda_{\rm u}^{\rm s}\over \tilde{k}_{0s}+u\,\epsilon_{\rm s}({\bf k})}\label{S5s5}
\eea
with $ \tilde{k}_{0s}= \tilde{k}_{0}-{\mu_{\rm Q}\over2}$ and $\Lambda_{\rm u}^{\rm s}={1\over2}\left(1+u\,\gamma^0{\boldsymbol{\gamma\cdot k}+m_{\rm s}\over \epsilon_{\rm s}({\bf k})}\right) $, the kaon part can be be further reduced to
\bea
\Omega_{\rm K}'=\sum_{n=1}{\beta_{\rm 2n}'\over n}{\cal K}^{2n},\ \beta_{\rm 2n}'\equiv{1\over V_4}{\rm Tr} \left(S_{\rm uu}S_{\rm 5s5}\right)^{n}.
\eea
By substituting the explicit expressions of $S_{\rm uu}$ and $S_{\rm 5s5}$ from Eqs. \eqref{Suu} and \eqref{S5s5}, respectively, the expansion coefficient $\beta_{\rm 2n}'$ can be calculated to any given order in principle. The quark propagators are diagonal in color space, so the trace over Dirac space is the most tough part in the calculation.

As we can see, the basic element in the trace terms is
\bea
S_{\rm uu}S_{\rm 5s5}=\sum_{t,u=\pm}{[\tilde{k}_{0s}+t\,\epsilon_{\rm l}({\bf k})]\Lambda_{\rm t} \Lambda_{\rm u}^{\rm s}
\over [\tilde{k}_0^2-(E_{\bf k}^{\rm -t})^2][\tilde{k}_{0s}+u\,\epsilon_{\rm s}({\bf k})]}.
\eea
To help evaluate higher order coefficients, it is useful to check the commuting relation between $\Lambda_{\rm t}$ and $\Lambda_{\rm u}^{\rm s}$ and we find
\bea
\!\!\!\!\!\! \Lambda_{\rm u}^{\rm s}\Lambda_{\rm t}\!=\!\Lambda_{\rm t}\Lambda_{\rm u}^{\rm s}\!+\!\delta\Lambda_{\rm u,t},\ \delta\Lambda_{\rm u,t}\!\equiv\!{1\over2}u\,t\,{\delta m\over \epsilon_{\rm l}\epsilon_{\rm s}}\boldsymbol{\gamma\cdot k}
\eea
with $\delta m\equiv m_{\rm s}-m_{\rm l}$. The term $\delta\Lambda_{\rm u,t}$ bears the following properties: 
\bea
&&\Lambda_{\rm t'}^{\rm (s)}\delta\Lambda_{\rm u,t}=\delta\Lambda_{\rm u,t}\Lambda_{\rm -t'}^{\rm (s)},\\
&&\delta\Lambda_{\rm u,t}\delta\Lambda_{\rm u',t'}=-{1\over4}u\,t\,u'\,t'\,f(k,m_{\rm l},m_{\rm s})
\eea
with $f(k,m_{\rm l},m_{\rm s})\equiv\left({k\delta m\over \epsilon_{\rm l}\epsilon_{\rm s}}\right)^2$. Now, we can evaluate the trace over Dirac space order by order by utilizing the properties of $\Lambda_{\rm t}^{\rm (s)}$ and recursions, that is,
\begin{widetext}
\bea
\alpha_{t,u}&\equiv&{\rm Tr}_{\rm D} \Lambda_{\rm t} \Lambda_{\rm u}^{\rm s}={\rm Tr}_{\rm D} \Lambda_{\rm u}^{\rm s}\Lambda_{\rm t}=1+t\,u\,{{\bf k}^2\!+\!m_{\rm s}m_{\rm l}\over \epsilon_{\rm l}\epsilon_{\rm s}};\\
\alpha_{t_1,u_1;t_2,u_2}&\equiv&{\rm Tr}_{\rm D} \Lambda_{\rm t_1}\Lambda_{\rm u_1}^{\rm s}\Lambda_{\rm t_2}\Lambda_{\rm u_2}^{\rm s}\!=\!{\rm Tr}_{\rm D} \Lambda_{\rm t_1}\left(\Lambda_{\rm t_2}\Lambda_{\rm u_1}^{\rm s}+\delta\Lambda_{\rm u_1,t_2}\right)\Lambda_{\rm u_2}^{\rm s}=\delta_{t_1,t_2}\delta_{u_1,u_2}\alpha_{t_1,u_1}+\delta^2\alpha_{t_1,u_1;t_2,u_2},\\
&&\delta^2\alpha_{t_1,u_1;t_2,u_2}\equiv{\rm Tr}_{\rm D} \Lambda_{\rm t_1}\delta\Lambda_{\rm u_1,t_2}\Lambda_{\rm u_2}^{\rm s}=-{1\over2}u_1\,t_1\,u_2\,t_2\,f(k,m_{\rm l},m_{\rm s});\nonumber\\
\alpha_{t_1,u_1;;t_3,u_3}&\equiv&{\rm Tr}_{\rm D} \Lambda_{\rm t_1}\Lambda_{\rm u_1}^{\rm s}\Lambda_{\rm t_2}\Lambda_{\rm u_2}^{\rm s}\Lambda_{\rm t_3}\Lambda_{\rm u_3}^{\rm s}=\delta_{t_1,t_2}\delta_{u_1,u_2}{\rm Tr}_{\rm D} \Lambda_{\rm t_1}\Lambda_{\rm u_1}^{\rm s}\Lambda_{\rm t_3}\Lambda_{\rm u_3}^{\rm s}\!+\!{\rm Tr}_{\rm D} \Lambda_{\rm t_1}\delta\Lambda_{\rm u_1,t_2}\Lambda_{\rm u_2}^{\rm s}\Lambda_{\rm t_3}\Lambda_{\rm u_3}^{\rm s}\nonumber\\
&=&\delta_{t_1,t_2}\delta_{u_1,u_2}\alpha_{t_1,u_1;t_3,u_3}\!+\!\delta_{t_1,-t_3}\delta_{u_2,u_3}\delta^2\alpha_{t_1,u_1;t_2,u_2}+{\rm Tr}_{\rm D} \Lambda_{\rm t_1}\delta\Lambda_{\rm u_1,t_2}\delta\Lambda_{\rm u_2,t_3}\Lambda_{\rm u_3}^{\rm s}\nonumber\\
&=&\delta_{t_1,t_2}\delta_{u_1,u_2}\alpha_{t_1,u_1;t_3,u_3}\!+\!\delta_{t_1,-t_3}\delta_{u_2,u_3}\delta^2\alpha_{t_1,u_1;t_2,u_2}+{1\over2}\alpha_{t_1,u_3}\delta^2\alpha_{t_2,u_1;t_3,u_2};\\
\alpha_{t_1,u_1;;t_4,u_4}&\equiv&{\rm Tr}_{\rm D} \Lambda_{\rm t_1}\Lambda_{\rm u_1}^{\rm s}\Lambda_{\rm t_2}\Lambda_{\rm u_2}^{\rm s}\Lambda_{\rm t_3}\Lambda_{\rm u_3}^{\rm s}\Lambda_{\rm t_4}\Lambda_{\rm u_4}^{\rm s}=\delta_{t_1,t_2}\delta_{u_1,u_2}\alpha_{t_2,u_2;;t_4,u_4}\!+\!{\rm Tr}_{\rm D} \Lambda_{\rm t_1}\delta\Lambda_{\rm u_1,t_2}\Lambda_{\rm u_2}^{\rm s}\Lambda_{\rm t_3}\Lambda_{\rm u_3}^{\rm s}\Lambda_{\rm t_4}\Lambda_{\rm u_4}^{\rm s}\nonumber\\
&=&\delta_{t_1,t_2}\delta_{u_1,u_2}\alpha_{t_2,u_2;;t_4,u_4}\!+\!\delta_{t_3,t_4}\delta_{u_3,u_4}\left(\delta_{t_1,-t_3}\delta_{u_2,u_3}\delta^2\alpha_{t_1,u_1;t_2,u_2}+{1\over2}\alpha_{t_1,u_3}\delta^2\alpha_{t_2,u_1;t_3,u_2}\right)\nonumber\\
&&+{1\over2}\alpha_{t_1,-u_2;-t_3,u_4}\delta^2\alpha_{t_2,u_1;t_4,u_3}.
\eea
We notice that the coefficients are all even functions of $\delta m$, and the leading nontrivial dependence is of quadratic except $\alpha_{t,u}$. Then, the expansion coefficients for the kaon part of thermodynamic potential follow as
\bea
\beta_{\rm 2n}'\equiv{1\over V_4}\sum_{k}\sum_{q_{\rm j}}\sum_{t_1,u_2;;t_{\rm n},u_{\rm n}=\pm}\alpha_{t_1,u_1;;t_n,u_n}\prod_{i=1,\dots,n}{[\tilde{k}_{0s}+t_{\rm i}\,\epsilon_{\rm l}({\bf k})]
\over [\tilde{k}_0^2-(E_{\bf k}^{\rm -t_{\rm i}})^2][\tilde{k}_{0s}+u_{\rm i}\,\epsilon_{\rm s}({\bf k})]},\label{Omgk}
\eea
\end{widetext}
where $q_{\rm j}\ (j=1,2,3)$ are the color charges of quarks and the conventions $\alpha_{t_1,u_1;;t_1,u_1}=\alpha_{t_1,u_1}$ and $\alpha_{t_1,u_1;;t_2,u_2}=\alpha_{t_1,u_1;t_2,u_2}$ should be understood. At finite temperature and chemical potential, we should work in Euclidean space by shifting $k_0$ to $i\omega_{\rm m}$ in $\beta_{\rm 2n}'$ and complete the summation over the Matsubara frequency $\omega_{\rm m}\equiv(2m+1)\pi T$ in order to facilitate numerical calculations, see Appendix.\ref{alpha} for more details. 

By the way, if we neglect $\delta m$ dependence in $\alpha_{t_1,u_1;;t_n,u_n}$, only the terms with $t_1=\cdots=t_n$ and $u_1=\cdots=u_n,\ (n=1,2,\dots)$ are nonzero and we simply have
\bea
\alpha_{t_1,u_1;;t_n,u_n}=\alpha_{t_1,u_1}\prod_{i=1}^n\delta_{t_1,t_i}\delta_{u_1,u_i}.
\eea
Then, the coefficients $\beta_{\rm 2n}'$ would be greatly simplified to
\bea
\beta_{\rm 2n}'\!=\!{1\over V_4}\!\sum_{k}\!\sum_{q_{\rm j}}\!\sum_{t,u=\pm}\!\!\!\alpha_{t,u}\!\!\left({\tilde{k}_{0s}+t\,\epsilon_{\rm l}({\bf k})
\over [\tilde{k}_0^2\!-\!(E_{\bf k}^{\rm -t})^2][\tilde{k}_{0s}\!+\!u\,\epsilon_{\rm s}({\bf k})]}\right)^n.\nonumber\\ \label{Omgk1}
\eea
Note that $\beta_{\rm 2}$ in Eq.\eqref{Omgk1} is exactly the same as that in Eq.\eqref{Omgk} as $\alpha_{t,u}$ does not depend on $\delta m$. With this approximation, the summation over $n$ can even be carried out analytically in $\Omega_{\rm K}'$ to arrive at
\bea
\Omega_{\rm K}'\!\!=\!\!{-1\over V_4}\!\!\sum_{k}\!\!\sum_{q_{\rm j}}\!\!\!\sum_{t,u=\pm}\!\!\!\!\alpha_{t,u}\!\ln\!\left\{\!1\!-\!{[\tilde{k}_{0s}+t\,\epsilon_{\rm l}({\bf k})]{\cal K}^{2}
\over [\tilde{k}_0^2\!-\!(E_{\bf k}^{\rm -t})^2][\tilde{k}_{0s}\!+\!u\,\epsilon_{\rm s}({\bf k})]}\!\right\}.\nonumber\\
\eea
We have checked numerically that such an approximation is very bad when $m_{\rm s}\gg m_{\rm l}$, so we will stick to the exact form \eqref{Omgk} in the following.

Eventually, the coupled gap equations follow directly from the definitions of condensates:
\bea
&&\sigma_{\rm s}\equiv\langle\bar{s}{s}\rangle={\partial\Omega_{\rm \pi s}'\over\partial m_{\rm s}},\
2\sigma_{\rm l}\equiv\langle\bar{u}{u}\rangle+\langle\bar{d}{d}\rangle={\partial\Omega_{\rm  \pi s}'\over\partial m_{\rm l}},\nonumber\\
&&2\Delta_\pi\equiv\langle\bar{u}i\gamma^5{d}\rangle+\langle\bar{d}i\gamma^5{u}\rangle={\partial\Omega_{\rm  \pi s}'\over\partial \Pi},\nonumber\\
&&2\Delta_{\rm K}\equiv\langle\bar{u}i\gamma^5{s}\rangle+\langle\bar{s}i\gamma^5{u}\rangle={\partial\Omega_{\rm K}'\over\partial {\cal K}}\label{conds}
\eea
and the minimal condition $\partial_{\rm L}[V(L,L)+\Omega_{\rm \pi s}]=0$ as~\cite{Cao:2021gfk}
\begin{widetext}
\bea
\sigma_{\rm s}
\!&=&\!-2N_c\int^\Lambda\!\!{\di^3k\over(2\pi)^3}{m_{\rm s}\over\epsilon_{\rm s}(k)}+2N_c\int\!\!{\di^3k\over(2\pi)^3}{m_{\rm s}\over\epsilon_{\rm s}(k)}\sum_{u=\pm}dV_1\left(\!L,u,\epsilon_{\rm s}(k),{-\mu_{\rm Q}\!+\!\mu_{\rm B}\over3}\!\right),\label{gaps}\\
2\sigma_{\rm l}\!&=&\!-2N_c\int^\Lambda\!\!{\di^3k\over(2\pi)^3}\sum_{t=\pm}{m_{\rm l}\over\epsilon_{\rm l}(k)}{\epsilon_{\rm l}(k)+t{\mu_{\rm Q}\over2}\over E_{\rm l}^{\rm t}(k)}+2N_c\int\!\!{\di^3k\over(2\pi)^3}\sum_{t,u=\pm}{m_{\rm l}\over\epsilon_{\rm l}(k)}{\epsilon_{\rm l}(k)+t{\mu_{\rm Q}\over2}\over E_{\rm l}^{\rm t}(k)}dV_1\left(\!L,u,E_{\rm l}^{\rm t}(k),{\mu_{\rm Q}\!+\!2\mu_{\rm B}\over6}\!\right),\label{gapl}\\
2\Delta_\pi\!&=&\!-2N_c\int^\Lambda\!\!{\di^3k\over(2\pi)^3}\sum_{t=\pm}{\Pi\over E_{\rm l}^{\rm t}(k)}+2N_c\int\!\!{\di^3k\over(2\pi)^3}\sum_{t,u=\pm}{\Pi\over E_{\rm l}^{\rm t}(k)}dV_1\left(\!L,u,E_{\rm l}^{\rm t}(k),{\mu_{\rm Q}\!+\!2\mu_{\rm B}\over6}\!\right),\label{gappi}\\
2\Delta_{\rm K}\!&=&2\sum_{n=1}{\beta_{\rm 2n}'}{\cal K}^{2n-1},\label{gapkaon}\\
&&\!\!\!\!\!\!\!\!\!\!\!\!\!\!T^4\left[-\left(3.51\!-{2.47\over\tilde{T}}\!+\!{15.2\over\tilde{T}^{2}}\right)L\!+\!{1.75\over\tilde{T}^{3}}{12L(1-L)^2\over1\!-\!6L^2\!+\!8L^3\!-\!3L^4}\right]=6T\!\int\!\!{\di^3k\over(2\pi)^3}\!\sum_{u=\pm} \left[\sum_{t=\pm}dV_2\left(L,u,E^{\rm t}(k),{\mu_{\rm Q}\!+\!2\mu_{\rm B}\over6}\right)\right.\nonumber\\
&&\left.+dV_2\left(\!L,u,\epsilon_{\rm s}(k),{-\mu_{\rm Q}\!+\!\mu_{\rm B}\over3}\!\right)\right].
\eea
Note that $\Delta_\pi=0$ and $\Delta_{\rm K}=0$ are trivial solutions of Eqs.\eqref{gappi} and \eqref{gapkaon}, respectively. As pion superfluidity is usually more favored than kaon superfluidity in this study, $\Delta_\pi$ (or $\Pi$) and $\Delta_{\rm K}$ (or ${\cal K}$) are actually true order parameters for $I_3$~\cite{Son:2000xc} and $I_8$ flavor symmetries, respectively. The total self-consistent thermodynamic potential can be found to be
\bea
\Omega_{\rm \pi KSF}&=&V(L,L)+\Omega_{\rm \pi s}'+\Omega_{\rm K}+2G(\sigma_{\rm s}^2+2\sigma_{\rm l}^2+2\Delta_\pi^2)-4K(\sigma_{\rm l}^2+\Delta_\pi^2)\sigma_{\rm s}
\eea
by utilizing the definitions of condensates and their relations to scalar and pseudoscalar masses, refer to Eqs.\eqref{conds} and \eqref{mDelta}. Here, the pion bilinear~\cite{Cao:2021gfk,Cao:2022fow} and full kaon parts of thermodynamic potential are, respectively,
\bea
\!\!\!\!\!\!\Omega_{\pi s}'&=&\!-\!2N_c\!\!\int^\Lambda\!\!\!\!{\di^3k\over(2\pi)^3}\!\!\left[\sum_{t=\pm}\!E_{\bf k}^{\rm t}\!+\!\epsilon_{\rm s}(k)\right]\!\!-\!2T\!\!\int\!\!\!{\di^3k\over(2\pi)^3}\!\!\sum_{\rm u=\pm}\!\!\left[\sum_{t=\pm}\!Fl\!\left(\!L,u,E_{\bf k}^{\rm t},{\mu_{\rm Q}\!+\!2\mu_{\rm B}\over6}\!\right)\!\!+\!Fl\!\left(\!L,u,\epsilon_{\rm s}(k),{-\mu_{\rm Q}\!+\!\mu_{\rm B}\over3}\!\right)\!\right],\\
&&\ \ \ \ \ \ \ \left\{Fl(L,u,x,y)=\log\left[1+3L\,e^{-{1\over T}\left(x-u\,y\right)}+3L\,e^{-{2\over T}\left(x-u\,y\right)}+e^{-{3\over T}\left(x-u\,y\right)}\right]\right\},\label{Omgpis}\\
\Omega_{\rm K}&\equiv& \Omega_{\rm K}'+(4G-2K\sigma_{\rm l})\Delta_{\rm K}^2=\sum_{n=1}{\beta_{\rm 2n}\over n}{\cal K}^{2n},\ \beta_{\rm 2n}=\beta_{\rm 2n}'+{\delta_{n,1}\over 4G-2K\sigma_{\rm l}}.
\eea
The contribution of $\Omega_{\rm K}$ is usually small compared to that of $\Omega_{\rm \pi s}'$ but is important for the evaluation of $\Delta_{\rm K}$. Following that, the entropy, electric charge and baryon densities of the system can be approximately evaluated by simply neglecting $\Omega_{\rm K}$ and we have~\cite{Cao:2021gfk}
\bea
\!\!\!\!\!\!s_{\rm \pi KSF}&=&2\!\!\int\!\!{\di^3k\over(2\pi)^3}\!\sum_{t,u=\pm}\left[Fl\left(\!L,u,E_{\rm l}^{\rm t}(k),{\mu_{\rm Q}\!+\!2\mu_{\rm B}\over6}\!\right)\!+\!{3\left(\!E_{\rm l}^{\rm t}(k)\!-\!u\,{\mu_{\rm Q}\!+\!2\mu_{\rm B}\over6}\!\right)\over T}dV_1\left(\!L,u,E_{\rm l}^{\rm t}(k),{\mu_{\rm Q}\!+\!2\mu_{\rm B}\over6}\!\right)\right]\nonumber\\
&&+2\!\!\int\!\!{\di^3k\over(2\pi)^3}\!\sum_{u=\pm}\left[Fl\left(\!L,u,\epsilon_{\rm s}(k),{-\mu_{\rm Q}\!+\!\mu_{\rm B}\over3}\!\right)\!+\!{3\left(\!E_{\rm l}^{\rm t}(k)\!-\!u\,{-\mu_{\rm Q}\!+\!\mu_{\rm B}\over3}\!\right)\over T}dV_1\left(\!L,u,\epsilon_{\rm s}(k),{-\mu_{\rm Q}\!+\!\mu_{\rm B}\over3}\!\right)\right]\nonumber\\
&&+T^3\left\{{1\over2}\left(4\times3.51-3\times{2.47\over\tilde{T}}+2\times{15.2\over\tilde{T}^{2}}\right)L^2+{1.75\over\tilde{T}^{3}}\ln\left[1-6L^2+8L^3-3L^4\right]\right\},\label{s3f}\\
\!\!\!\!\!\!n_{\rm Q}^{\rm \pi KSF}&=&N_c\!\!\int^\Lambda\!\!{\di^3k\over(2\pi)^3}\!\sum_{t=\pm}t{\epsilon_{\rm l}(k)\!+\!t{\mu_{\rm Q}\over2}\over E_{\rm l}^{\rm t}(k)}-\!3\!\!\int\!\!{\di^3k\over(2\pi)^3}\sum_{t,u=\pm}t{\epsilon_{\rm l}(k)\!+\!t{\mu_{\rm Q}\over2}\over E_{\rm l}^{\rm t}(k)}dV_1\left(\!L,u,E_{\rm l}^{\rm t}(k),{\mu_{\rm Q}\!+\!2\mu_{\rm B}\over6}\!\right)\nonumber\\
&&+\int\!\!{\di^3k\over(2\pi)^3}\!\sum_{t,u=\pm}u\,dV_1\left(\!L,u,E_{\rm l}^{\rm t}(k),{\mu_{\rm Q}\!+\!2\mu_{\rm B}\over6}\!\right)-2\int{\di^3k\over(2\pi)^3}\!\sum_{t,u=\pm}u\,dV_1\left(\!L,u,\epsilon_{\rm s}(k),{-\mu_{\rm Q}\!+\!\mu_{\rm B}\over3}\!\right),\label{nQ3f}\\
\!\!\!\!\!\!n_{\rm B}^{\rm \pi KSF}&=&2\int{\di^3k\over(2\pi)^3}\!\sum_{t,u=\pm}u\,dV_1\left(\!L,u,E_{\rm l}^{\rm t}(k),{\mu_{\rm Q}\!+\!2\mu_{\rm B}\over6}\!\right)+2\int{\di^3k\over(2\pi)^3}\!\sum_{t,u=\pm}u\,dV_1\left(\!L,u,\epsilon_{\rm s}(k),{-\mu_{\rm Q}\!+\!\mu_{\rm B}\over3}\!\right).\label{nB3f}
\eea
\end{widetext}

\subsubsection{The electroweak interaction sector}\label{EW}
In free gas approximation, the thermodynamic potentials for the QEWD sector are~\cite{Kapusta2006}
\bea
\Omega_\gamma&=&2T\int{\di^3k\over(2\pi)^3}\log\left(1-e^{- k/T}\right),\\
\Omega_{\rm l}&=&-T\!\!\sum^{\rm i=e,\mu,\tau}_{u=\pm}\!\int\!\!{\di^3k\over(2\pi)^3}\!\left\{2\log\left[1+e^{- \left(\epsilon_{\rm i}(k)-u\,(-\mu_{\rm Q}+\mu_{\rm i})\right)/T}\right]\right.\nonumber\\
&&\left.+\log\left[1+e^{- (k-u\,\mu_{\rm i})/T}\right]\right\},\label{Omegal}
\eea
and the corresponding entropy, electric charge and lepton flavor densities are, respectively,
\bea
s_\gamma&=&2\int{\di^3k\over(2\pi)^3}\left[-\log\left(1-e^{- k/T}\right)+{k/T\over e^{k/T}-1}\right],\\
s_{\rm l}&=&\sum^{\rm i=e,\mu,\tau}_{u=\pm}\int{\di^3k\over(2\pi)^3}\left\{2\log\left[1+e^{- \left(\epsilon_{\rm i}(k)-u\,(-\mu_{\rm Q}+\mu_{\rm i})\right)/T}\right]\right.\nonumber\\
&&+\log\left[1+e^{- (k-u\,\mu_{\rm i})/T}\right]\!+\!{2\left(\epsilon_{\rm i}(k)\!-\!u\,(\!-\mu_{\rm Q}\!+\!\mu_{\rm i})\right)/T\over1+e^{\left(\epsilon_{\rm i}(k)-u\,(-\mu_{\rm Q}+\mu_{\rm i})\right)/T}}\nonumber\\
&&\left.+{(k-u\,\mu_{\rm i})/T\over1+e^{(k-u\,\mu_{\rm i})/T}}\right\},
\eea
\bea
n_{\rm Q}^{\rm l}&=&2T\sum^{\rm i=e,\mu,\tau}_{u=\pm}\int{\di^3k\over(2\pi)^3}{-u\over1+e^{\left(\epsilon_{\rm i}(k)-u\,(-\mu_{\rm Q}+\mu_{\rm i})\right)/T}},\\
n_{\rm i}&=&-{\partial \Omega_{\rm l}\over \partial \mu_{\rm i}}=T\sum_{u=\pm}\int{\di^3k\over(2\pi)^3}\left[{2u\over1+e^{\left(\epsilon_{\rm i}(k)-u\,(-\mu_{\rm Q}+\mu_{\rm i})\right)/T}}\right.\nonumber\\
&&\qquad\qquad\ \ \ \left.+{u\over1+e^{(k-u\,\mu_{\rm i})/T}}\right],\ i=e,\mu,\tau.
\eea

So, in the pion or kaon superfluidity, the total thermodynamic potential, entropy, electric charge and lepton densities are, respectively, 
\bea
\Omega&=&\Omega_\gamma+\Omega_{\rm l}+\Omega_{\rm \pi KSF},\,
s=s_\gamma\!+\!s_{\rm l}\!+\!s_{\rm \pi KSF},\nonumber\\
n_{\rm Q}&=&n_{\rm Q}^{\rm l}\!+\!n_{\rm Q}^{\rm \pi KSF},\ \ \ \ \ \ \ \,
n_{\rm l}=\!\sum_{\rm i=e,\mu,\tau} \!\! n_{\rm i}
\eea
after combining the QEWD and QCD sectors together. According to the conventions, we define the following reduced quantities:
\bea
b=n_{\rm B}^{\rm \pi KSF}/s,\ l=n_{\rm l}/s,\ l_{\rm i}=n_{\rm i}/s.
\eea

\section{Numerical results for the case without magnetic field}\label{numerical}
To carry out numerical calculations, we fix the electron and muon masses from the Particle Data Group as $m_{\rm e}=0.53\,{\rm MeV}$ and $m_\mu=113\,{\rm MeV}$ and suppress the contribution of heavy $\tau$ lepton. The model parameters for the strong interaction sector are fixed as the following~\cite{Zhuang:1994dw,Rehberg:1995kh}
\bea
&& m_{\rm 0l}\!=\!5.5\,{\rm MeV},\, m_{\rm 0s}\!=\!140.7\,{\rm MeV},\, \Lambda\!=\!602.3\,{\rm MeV},\nonumber\\ 
&&G\Lambda^2\!=\!1.835,\, K\Lambda^5\!=\!12.36.
\eea
In this work, we only consider the case without primordial magnetic field, and simply set $l=-0.012$ and $l_{\rm e}=0$ as the results would not be affected qualitatively. As we have mentioned, in the case of full chemical balance with $\mu_{\rm d}=\mu_{\rm s}$, we expect $|{\cal K}|\leq |\Pi|$ since $\pi^\pm$ mass is much smaller than $K^\pm$ mass. So we maily focus on the pion superfluidity regime and check if kaon superfluidity could coexist with that in the following. 

\begin{figure}[!htb]
	\begin{center}
	\includegraphics[width=8cm]{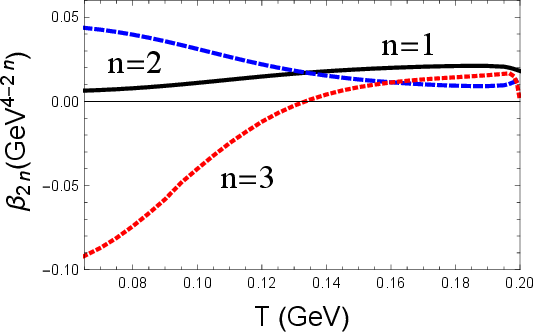}
	\includegraphics[width=8cm]{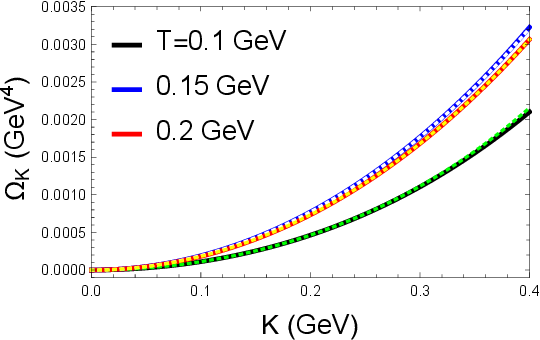}
		\caption{Upper: the coefficients $\beta_{\rm 2n} (n=1,2,3)$ as functions of temperature $T$ for lepton flavor asymmetry $l_{\rm e}+l_\mu=-0.3$, lower: the corresponding kaon part of thermodynamic potential $\Omega_{\rm K}$ as a function of ${\cal K}$ for $T=0.1, 0.15$ and $0.2~{\rm GeV}$ with solid lines to $o({\cal K}^6)$ and dotted lines to $o({\cal K}^4)$.}\label{beta_T}
	\end{center}
\end{figure}

For the lepton flavor asymmetry $l_{\rm e}+l_\mu=-0.3$, the coefficients $\beta_{\rm 2n} (n=1,2,3)$ are illustrated as functions of temperature in the upper panel of Fig.~\ref{beta_T}, and the corresponding kaon part of thermodynamic potential $\Omega_{\rm K}$ is shown for different temperatures in the lower panel of Fig.~\ref{beta_T}. Note: so many terms are involved in the full evaluation of $\beta_6$ that we only consider $\beta_{6}'^{\rm V}$ for demonstration here and in the following; the largest value of $\Pi$ is about $0.36~{\rm GeV}$ in this case, so the range ${\cal K}\in[0,0.4]~{\rm GeV}$ is large enough to cover all possible minima of $\Omega_{\rm K}$. According to the upper panel, the signs of $\beta_2$ and $\beta_4$ keep positive up to $T_{\rm c}=0.21~{\rm GeV}$, while the sign of $\beta_6$ changes from negative to positive with increasing temperature. For all the chosen temperatures, it is easy to check from the lower panel that the quartic approximation up to $o({\cal K}^4)$ is very precise compared to that up to $o({\cal K}^6)$.  Since there is no extra minimum of $\Omega_{\rm K}$ in the considered range except at ${\cal K}=0$, the kaon superfluidity is not favored in this case. The medium part $\beta_{6}'^{\rm M}$ would become important with increasing temperature but should be at most of the same order as $\beta_{6}'^{\rm V}$, therefore the quartic approximation is still very good and the quantitative results remain. In that sense, though the expressions of $\beta_{\rm 2n}$ are very complicated for $n\geq3$, we can safely stick to the quartic Ginzburg-Landau approximation.

\begin{figure}[!htb]
	\begin{center}
	\includegraphics[width=8cm]{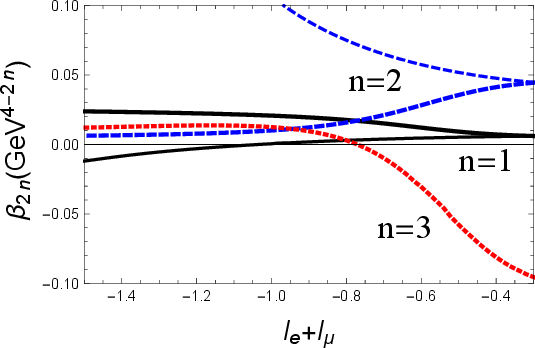}
	\includegraphics[width=8cm]{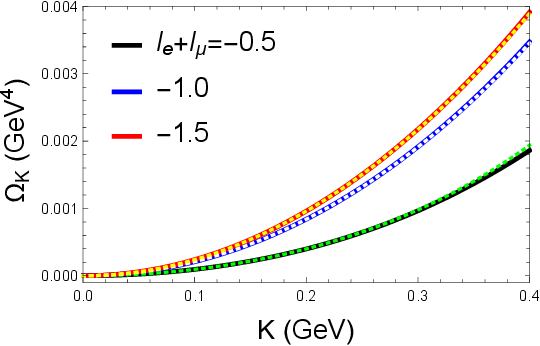}
		\caption{Upper: the coefficients $\beta_{\rm 2n} (n=1,2,3)$ as functions of lepton flavor asymmetry $l_{\rm e}+l_\mu$ at temperature $T=0.06~{\rm GeV}$ in pion superfluidity (thick lines) and chiral phases (thin lines), lower: the corresponding kaon part of thermodynamic potential $\Omega_{\rm K}$ as a function of ${\cal K}$ in the pion superfluid phase for $l_{\rm e}+l_\mu=-0.5, -1$ and $-1.5$ with solid lines to $o({\cal K}^6)$ and dotted lines to $o({\cal K}^4)$. }\label{beta_l}
	\end{center}
\end{figure}

In Fig.~\ref{beta_T}, the charge chemical potential $\mu_{\rm Q}$ is never found to satisfy $|\mu_{\rm Q}|>m_{\rm K}^{\rm V}\ (\sim0.5~{\rm GeV})$, so we are not surprised to find that kaon superfluidity is not favored. How about the case when we achieve $|\mu_{\rm Q}|>m_{\rm K}^{\rm V}$ by increasing $|l_{\rm e}+l_\mu|$? For a fixed temperature, $T=0.06~{\rm GeV}$, the coefficients $\beta_{\rm 2n} (n=1,2,3)$ are illustrated as functions of lepton flavor asymmetry $l_{\rm e}+l_\mu$ in the upper panel of Fig.~\ref{beta_l}, and the corresponding $\Omega_{\rm K}$ is shown for different $l_{\rm e}+l_\mu$ in the lower panel of Fig.~\ref{beta_l}. Similar to Fig.~\ref{beta_T}, the quartic approximation up to $o({\cal K}^4)$ is very precise compared to that up to $o({\cal K}^6)$ according to the lower panel. According to the upper panel, $\beta_2$ decreases with increasing $|l_{\rm e}+l_\mu|$ and becomes negative when $|\mu_{\rm Q}|>m_{\rm K}^{\rm V}$ in the chiral phases; however, $\beta_2$ increases with increasing $|l_{\rm e}+l_\mu|$ and keeps positive in the pion superfluidity. As $\beta_4$ is positive for both cases, we conclude that kaon superfluidity is possible in the background of chiral phases but would get killed by pion superfluidity. Combine the results in Fig.~\ref{beta_T} and Fig.~\ref{beta_l}, both $\beta_2$ and $\beta_4$ keep positive with increasing $T$ or $|l_{\rm e}+l_\mu|$, thus kaon superfluidity is not possible to coexist with pion superfluidity in the case of chemical balance.

\begin{figure}[!htb]
	\begin{center}
	\includegraphics[width=8cm]{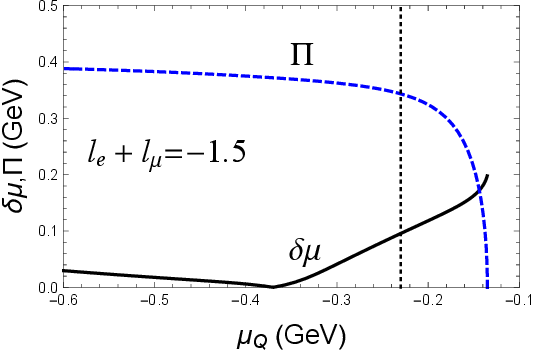}
		\caption{For the lepton flavor asymmetry $l_{\rm e}+l_\mu=-1.5$, the mismatch $\delta\mu$ and pion condensate $\Pi$ as functions of $\mu_{\rm Q}$ by evolving the temperature in the range $(15,64)~{\rm MeV}$. The vertical line corresponds to $\mu_{\rm Q}=-0.23~{\rm GeV}$.}\label{deltamu}
	\end{center}
\end{figure}

According to Ref.~\cite{He:2006tn}, the pion superfluidity could be in Sarma phase for $m_\pi^{\rm V}<|\mu_{\rm Q}|<0.23~{\rm GeV}$ and in Larkin-Ovchinnikov-Fudde-Ferrell (LOFF) phase for $|\mu_{\rm Q}|>0.23~{\rm GeV}$ at small baryon density and zero temperature. Besides, by following the discussions on two-flavor color superconductor~\cite{Cao:2015rea},  the LO phase with an antipodal-wave form $\Pi(z)=\Pi \cos(q z)$ is expected to be favored in the mismatch range $\delta\mu\equiv |(\mu_{\rm Q}+2\mu_{\rm B})/6|\in(0.7,0.8)\Pi_0$, where $\Pi_0$ is the pion condensate at $\delta\mu=0$. So, $\Pi(z)$ could vanish at spatial points $z_{\rm n}=(2n+1){\pi\over2}q^{-1},\ n=0,\pm1,\dots$ in the LOFF phase and kaon condensation would be nonzero there if $|\mu_{\rm Q}|>m_{\rm K}$, see the discussions on the upper panel of Fig.~\ref{beta_l}. By choosing a large enough lepton flavor asymmetry, $l_{\rm e}+l_\mu=-1.5$, the mismatch and pion condensate are shown together as functions of $\mu_{\rm Q}$ in Fig.~\ref{deltamu} by evolving the temperature in the range $(15,64)~{\rm MeV}$. As $\Pi\leq\Pi_0$ for a given $\mu_{\rm Q}$, we never find $\delta\mu$ to be in the range $(0.7,0.8)\Pi_0$ for $|\mu_{\rm Q}|>0.23~{\rm GeV}$, hence the LOFF phase cannot be realized and the inhomogeneous kaon superfluidity is disfavored.

\begin{figure}[!htb]
	\begin{center}
	\includegraphics[width=8cm]{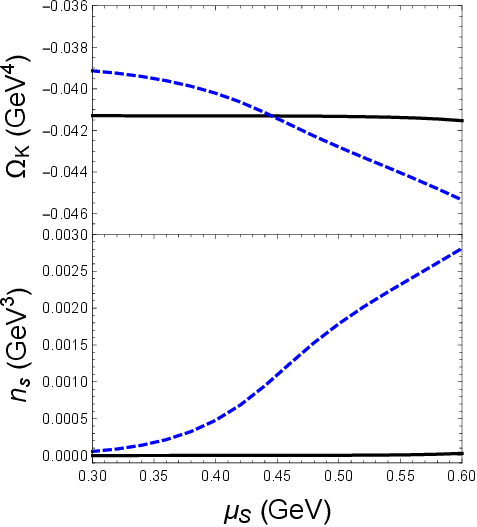}
		\caption{For the temperature $T=0.06~{\rm GeV}$ and lepton flavor asymmetry $l_{\rm e}+l_\mu=-1.2$, the kaon part of thermodynamic potential $\Omega_{\rm K}$ (upper panel) and $s$ quark density $n_{\rm s}$ (lower panel) as functions of strangeness chemical potential $\mu_{\rm S}$ in the pion superfluidity (black solid lines) and chiral phases (blue dashed lines).}\label{muS}
	\end{center}
\end{figure}

In order to check how hard can kaon superfluidity be realized in the early universe, we introduce the strangeness chemical potential $\mu_{\rm S}$ to $s$ quark and then the full chemical potential changes to $\mu_{\rm s}=(-\mu_{\rm Q}+\mu_{\rm B})/3+\mu_{\rm S}$. As we can tell, $\mu_{\rm S}$ now plays a role of mismatch between $s$ and $d$ quarks, so $s$ quark density increases with $\mu_{\rm S}$ and $d$ quark density would get suppressed since the total baryon density is small. Eventually, it can be realized that kaon superfluidity, instead of pion superfluidity, is the ground state of the system. For $T=0.06~{\rm GeV}$ and $l_{\rm e}+l_\mu=-1.2$, the kaon part of thermodynamic potential $\Omega_{\rm K}$ and $s$ quark density $n_{\rm s}$ are illustrated as functions of $\mu_{\rm S}$ in Fig.~\ref{muS}. We can easily identify a first-order transition from pion superfluidity to chiral phases around $\mu_{\rm S}=0.45~{\rm GeV}$ in the upper panel, but the latter is unstable to kaon superfluidity as $|\mu_{\rm Q}|>m_{\rm K}$ there. According to the lower panel, $n_{\rm s}$ is required to be three orders larger than $n_{\rm d}$ in order to realize kaon superfluidity. If the original mismatch between $s$ and $d$ quarks is generated in the electroweak epoch ($10^{-11}s$ after big bang), it is impossible to obtain such a large mismatch in the QCD epoch ($10^{-6}s$ after big bang) since the lifetime of $s$ quark is $10^{-8}s$.

\section{Summary}\label{summary}
In this work, the possibility of kaon superfluidity is checked for the QCD epoch of the early universe by considering the case with a vanishing primordial magnetic field and a large lepton flavor asymmetry. We work within the three-flavor Polyakov--Nambu--Jona-Lasinio model and adopt the Ginzburg-Landau approximation to develop a formalism for the study of kaon condensation within pion superfluidity. Actually, numerical calculations show that quartic Ginzburg-Landau approximation is precise enough to determine the value of kaon condensates. In the case of full chemical balances among elementary particles, kaon superfluidity could be stable compared to the chiral phases with only $\sigma$ condensations, but it would get killed by the true ground state -- homogeneous pion superfluidity. As baryon density is small in the early universe, we can only find Sarma gapless phase but not LOFF phase for the pion superfluidity at low temperature, so inhomogeneous kaon superfluidity seems impossible either. If we relax the chemical balance between $s$ and $d$ quark by introducing the strangeness chemical potential $\mu_{\rm S}$, kaon superfluidity could be theoretically realized by increasing $\mu_{\rm S}$ whence pion superfluidity becomes disfavored. But the corresponding $s$ quark density is so larger than $d$ quark density that we do not think that is realistic whence the lifetime of $s$ quark is self-consistently taken into account. By the way, a primordial magnetic field is not expected to cure the issue since $s$ and $d$ quarks respond the same to it. In a word, kaon superfluidity is impossible in the early universe. 

\section*{Acknowledgement}
G. C. is supported by the Natural Science Foundation of Guangdong Province with Grant No. 2024A1515011225.

\appendix
\begin{widetext}
\section{Expansion coefficients}\label{alpha}
At finite temperature and chemical potential, the coefficient $\beta_{\rm 2n}'$ in Eq.\eqref{Omgk} becomes
\bea
\beta_{\rm 2n}'\equiv -\int{\di^3{\bf k}\over(2\pi)^3}T\sum_{\rm m}\sum_{q_{\rm j}}\sum_{t_1,u_2;;t_{\rm n},u_{\rm n}=\pm}\alpha_{t_1,u_1;;t_n,u_n}\prod_{i=1,\dots,n}{[i\tilde{\omega}_{\rm m,j}-{\mu_{\rm Q}\over2}+t_{\rm i}\,\epsilon_{\rm l}({\bf k})]
\over [\tilde{\omega}_{\rm m,j}^2+(E_{\bf k}^{\rm -t_{\rm i}})^2][i\tilde{\omega}_{\rm m,j}-{\mu_{\rm Q}\over2}+u_{\rm i}\,\epsilon_{\rm s}({\bf k})]},
\eea
where $i\tilde{\omega}_{\rm m,j}=i\omega_{\rm m}+iq_{\rm j}T+{\mu_{\rm Q}+2\mu_{\rm B}\over6}$ with the fermion Matsubara frequency $\omega_{\rm m}\equiv(2m+1)\pi T$. To facilitate numerical calculations, we are going to complete the summations over $\omega_{\rm m}$ and $q_{\rm j}$ analytically, that is, explicitly evaluating the general term
\bea
\xi_{\rm 2n}\equiv -T\sum_{\rm m}\sum_{q_{\rm j}}\prod_{i=1,\dots,n}{[i\tilde{\omega}_{\rm m,j}-{\mu_{\rm Q}\over2}+t_{\rm i}\,\epsilon_{\rm l}({\bf k})]
\over [\tilde{\omega}_{\rm m,j}^2+(E_{\bf k}^{\rm -t_{\rm i}})^2][i\tilde{\omega}_{\rm m,j}-{\mu_{\rm Q}\over2}+u_{\rm i}\,\epsilon_{\rm s}({\bf k})]}.
\eea
By introducing contour integral around the imaginary axis, the summation over $\omega_{\rm m}$ can be alternatively presented as 
\bea
\xi_{\rm 2n}\equiv\left(\int_{-i\infty+\varepsilon}^{i\infty+\varepsilon}+\int_{i\infty-\varepsilon}^{-i\infty-\varepsilon}\right){\di k_0\over4\pi i}\tanh{k_0\over2T}
\sum_{q_{\rm j}}\prod_{i=1,\dots,n}{[\tilde{k}_{\rm 0j}-{\mu_{\rm Q}\over2}+t_{\rm i}\,\epsilon_{\rm l}({\bf k})]\over [\tilde{k}_{\rm 0j}^2-(E_{\bf k}^{\rm -t_{\rm i}})^2][\tilde{k}_{\rm 0j}-{\mu_{\rm Q}\over2}+u_{\rm i}\,\epsilon_{\rm s}({\bf k})]}
\eea
with $\tilde{k}_{\rm 0j}=k_0+iq_{\rm j}T+{\mu_{\rm Q}+2\mu_{\rm B}\over6}$. Then, by including the vanishing integral over lines with $|k_0|=\infty$, the contour integral can be separated into two clockwise integral loops for $\Re (k_0)>0$ and $\Re (k_0)<0$. To carry out the complex variable integral, it is convenient to rewrite the coefficient as
\bea
\xi_{\rm 2n}\equiv\left(\int_{-i\infty+\varepsilon}^{i\infty+\varepsilon}+\int_{i\infty-\varepsilon}^{-i\infty-\varepsilon}\right){\di k_0\over4\pi i}
\sum_{q_{\rm j}}{[\tilde{k}_{\rm 0j}-{\mu_{\rm Q}\over2}+\epsilon_{\rm l}({\bf k})]^{n+T_{\rm n}\over2}[\tilde{k}_{\rm 0j}-{\mu_{\rm Q}\over2}-\epsilon_{\rm l}({\bf k})]^{n-T_{\rm n}\over2}\tanh{k_0\over2T}\over [\tilde{k}_{\rm 0j}^2-(E_{\bf k}^{\rm -})^2]^{n+T_{\rm n}\over2} [\tilde{k}_{\rm 0j}^2-(E_{\bf k}^{\rm +})^2]^{n-T_{\rm n}\over2}[\tilde{k}_{\rm 0j}-{\mu_{\rm Q}\over2}+\epsilon_{\rm s}({\bf k})]^{n+U_{\rm n}\over2}[\tilde{k}_{\rm 0j}-{\mu_{\rm Q}\over2}-\epsilon_{\rm s}({\bf k})]^{n-U_{\rm n}\over2}}
\eea
by collecting all the terms with same $t_{\rm i}$ or $u_{\rm i}$ together. Here, we have defined $T_{\rm n}\equiv\sum_{i=1,\dots,n}t_{\rm i}$ and $U_{\rm n}\equiv\sum_{i=1,\dots,n}u_{\rm i}$ which are both in the range $[-n,n]$. Then, ${n\pm T_{\rm n}\over2},{n\pm U_{\rm n}\over2}\in[0,n]$.

Eventually, we apply the theorem of complex variable integral to find
\bea
\xi_{\rm 2n}&\equiv&-\sum_{v=\pm}\sum_{q_{\rm j}}\left\{{\partial^{{n+T_{\rm n}\over2}-1}_x\over\left({n+T_{\rm n}\over2}-1\right)!}{[x-{\mu_{\rm Q}\over2}+\epsilon_{\rm l}({\bf k})]^{n+T_{\rm n}\over2}[x-{\mu_{\rm Q}\over2}-\epsilon_{\rm l}({\bf k})]^{n-T_{\rm n}\over2}\tanh{x-(iq_{\rm j}T+{\mu_{\rm Q}+2\mu_{\rm B}\over6})\over2T}\over 2(x+v\,E_{\bf k}^{\rm -})^{n+T_{\rm n}\over2} [x^2-(E_{\bf k}^{\rm +})^2]^{n-T_{\rm n}\over2}[x-{\mu_{\rm Q}\over2}+\epsilon_{\rm s}({\bf k})]^{n+U_{\rm n}\over2}[x-{\mu_{\rm Q}\over2}-\epsilon_{\rm s}({\bf k})]^{n-U_{\rm n}\over2}}\Big|_{x=v\,E_{\bf k}^{\rm -}}\right.\nonumber\\
&&+{\partial^{{n-T_{\rm n}\over2}-1}_x\over\left({n-T_{\rm n}\over2}-1\right)!}{[x-{\mu_{\rm Q}\over2}+\epsilon_{\rm l}({\bf k})]^{n+T_{\rm n}\over2}[x-{\mu_{\rm Q}\over2}-\epsilon_{\rm l}({\bf k})]^{n-T_{\rm n}\over2}\tanh{x-(iq_{\rm j}T+{\mu_{\rm Q}+2\mu_{\rm B}\over6})\over2T}\over 2(x+v\,E_{\bf k}^{\rm +})^{n-T_{\rm n}\over2} [x^2-(E_{\bf k}^{\rm -})^2]^{n+T_{\rm n}\over2}[x-{\mu_{\rm Q}\over2}+\epsilon_{\rm s}({\bf k})]^{n+U_{\rm n}\over2}[x-{\mu_{\rm Q}\over2}-\epsilon_{\rm s}({\bf k})]^{n-U_{\rm n}\over2}}\Big|_{x=v\,E_{\bf k}^{\rm +}}\nonumber\\
&&+{\partial^{{n+U_{\rm n}\over2}-1}_x\over\left({n+U_{\rm n}\over2}-1\right)!}{[x+\epsilon_{\rm l}({\bf k})]^{n+T_{\rm n}\over2}[x-\epsilon_{\rm l}({\bf k})]^{n-T_{\rm n}\over2}\tanh{x-(iq_{\rm j}T+{-\mu_{\rm Q}+\mu_{\rm B}\over3})\over2T}\over 4 [(x+{\mu_{\rm Q}\over2})^2-(E_{\bf k}^{\rm -})^2]^{n-T_{\rm n}\over2} [(x+{\mu_{\rm Q}\over2})^2-(E_{\bf k}^{\rm +})^2]^{n-T_{\rm n}\over2}[x-\epsilon_{\rm s}({\bf k})]^{n-U_{\rm n}\over2}}\Big|_{x=-\epsilon_{\rm s}({\bf k})}\nonumber\\
&&\left.+{\partial^{{n-U_{\rm n}\over2}-1}_x\over\left({n-U_{\rm n}\over2}-1\right)!}{[x+\epsilon_{\rm l}({\bf k})]^{n+T_{\rm n}\over2}[x-\epsilon_{\rm l}({\bf k})]^{n-T_{\rm n}\over2}\tanh{x-(iq_{\rm j}T+{-\mu_{\rm Q}+\mu_{\rm B}\over3})\over2T}\over 4 [(x+{\mu_{\rm Q}\over2})^2-(E_{\bf k}^{\rm -})^2]^{n-T_{\rm n}\over2} [(x+{\mu_{\rm Q}\over2})^2-(E_{\bf k}^{\rm +})^2]^{n-T_{\rm n}\over2}[x+\epsilon_{\rm s}({\bf k})]^{n+U_{\rm n}\over2}}\Big|_{x=\epsilon_{\rm s}({\bf k})}\right\},
\eea
where the conventions $\partial^{0}_x/0!=1$ and $\partial^{-1}_x/(-1)!=0$ should be understood. The summation over $q_{\rm j}$ can be completed ahead by utilizing the following property
\bea
\sum_{q_{\rm j}}\tanh{x-i\,q_{\rm j}T\over2T}&=&\sum_{q_{\rm j}}\tanh{x-i\,q_{\rm j}T\over2T}=\sum_{q_{\rm j}} {1-e^{-x/T+i\,q_{\rm j}}\over 1+e^{-x/T+i\,q_{\rm j}}}=3+2T\partial_x\sum_{q_{\rm j}}\ln\left(1+e^{-x/T+i\,q_{\rm j}}\right)\nonumber\\
&=&3+2T\partial_x\ln\left(1+3L\,e^{-x\over T}+3L^*e^{-2x\over T}+e^{-3x\over T}\right)=3-6g(x,T,L,L^*)=-3+6g^*(-x,T,L,L^*),\nonumber\\
&&g(x,T,L,L^*)\equiv{L\,e^{-x\over T}+2L^*e^{-2x\over T}+e^{-3x\over T}\over 1+3L\,e^{-x\over T}+3L^*e^{-2x\over T}+e^{-3x\over T}}
\eea
with $\lim_{x\rightarrow\infty}g(x,T,L,L^*)=\lim_{x\rightarrow-\infty}g^*(-x,T,L,L^*)=0$. By following the assumption $L=L^*$, we have $g(x,T,L,L)=g^*(x,T,L,L)$ and then the function be simply rewritten as $g(x,T,L)$ instead. With the help of $g(x,T,L)$, the coefficient can be separated into vacuum and medium parts, that is, $\xi_{\rm 2n}\equiv\xi_{\rm 2n}^{\rm V}+\xi_{\rm 2n}^{\rm M}$ with
\bea
\xi_{\rm 2n}^{\rm V}&\equiv&-N_{\rm c}\sum_{v=\pm}\left\{{\partial^{{n+T_{\rm n}\over2}-1}_x\over\left({n+T_{\rm n}\over2}-1\right)!}{v[x-{\mu_{\rm Q}\over2}+\epsilon_{\rm l}({\bf k})]^{n+T_{\rm n}\over2}[x-{\mu_{\rm Q}\over2}-\epsilon_{\rm l}({\bf k})]^{n-T_{\rm n}\over2}\over 2(x+v\,E_{\bf k}^{\rm -})^{n+T_{\rm n}\over2} [x^2-(E_{\bf k}^{\rm +})^2]^{n-T_{\rm n}\over2}[x-{\mu_{\rm Q}\over2}+\epsilon_{\rm s}({\bf k})]^{n+U_{\rm n}\over2}[x-{\mu_{\rm Q}\over2}-\epsilon_{\rm s}({\bf k})]^{n-U_{\rm n}\over2}}\Big|_{x=v\,E_{\bf k}^{\rm -}}\right.\nonumber\\
&&+{\partial^{{n-T_{\rm n}\over2}-1}_x\over\left({n-T_{\rm n}\over2}-1\right)!}{v[x-{\mu_{\rm Q}\over2}+\epsilon_{\rm l}({\bf k})]^{n+T_{\rm n}\over2}[x-{\mu_{\rm Q}\over2}-\epsilon_{\rm l}({\bf k})]^{n-T_{\rm n}\over2}\over 2(x+v\,E_{\bf k}^{\rm +})^{n-T_{\rm n}\over2} [x^2-(E_{\bf k}^{\rm -})^2]^{n+T_{\rm n}\over2}[x-{\mu_{\rm Q}\over2}+\epsilon_{\rm s}({\bf k})]^{n+U_{\rm n}\over2}[x-{\mu_{\rm Q}\over2}-\epsilon_{\rm s}({\bf k})]^{n-U_{\rm n}\over2}}\Big|_{x=v\,E_{\bf k}^{\rm +}}\nonumber\\
&&-{\partial^{{n+U_{\rm n}\over2}-1}_x\over\left({n+U_{\rm n}\over2}-1\right)!}{[x+\epsilon_{\rm l}({\bf k})]^{n+T_{\rm n}\over2}[x-\epsilon_{\rm l}({\bf k})]^{n-T_{\rm n}\over2}\over 4 [(x+{\mu_{\rm Q}\over2})^2-(E_{\bf k}^{\rm -})^2]^{n-T_{\rm n}\over2} [(x+{\mu_{\rm Q}\over2})^2-(E_{\bf k}^{\rm +})^2]^{n-T_{\rm n}\over2}[x-\epsilon_{\rm s}({\bf k})]^{n-U_{\rm n}\over2}}\Big|_{x=-\epsilon_{\rm s}({\bf k})}\nonumber\\
&&\left.+{\partial^{{n-U_{\rm n}\over2}-1}_x\over\left({n-U_{\rm n}\over2}-1\right)!}{[x+\epsilon_{\rm l}({\bf k})]^{n+T_{\rm n}\over2}[x-\epsilon_{\rm l}({\bf k})]^{n-T_{\rm n}\over2}\over 4 [(x+{\mu_{\rm Q}\over2})^2-(E_{\bf k}^{\rm -})^2]^{n-T_{\rm n}\over2} [(x+{\mu_{\rm Q}\over2})^2-(E_{\bf k}^{\rm +})^2]^{n-T_{\rm n}\over2}[x+\epsilon_{\rm s}({\bf k})]^{n+U_{\rm n}\over2}}\Big|_{x=\epsilon_{\rm s}({\bf k})}\right\},\\
\xi_{\rm 2n}^{\rm M}&\equiv&6\sum_{v=\pm}\left\{{\partial^{{n+T_{\rm n}\over2}-1}_x\over\left({n+T_{\rm n}\over2}-1\right)!}{v[x-{\mu_{\rm Q}\over2}+\epsilon_{\rm l}({\bf k})]^{n+T_{\rm n}\over2}[x-{\mu_{\rm Q}\over2}-\epsilon_{\rm l}({\bf k})]^{n-T_{\rm n}\over2}g\left(vx-v{\mu_{\rm Q}+2\mu_{\rm B}\over6},T,L\right)\over 2(x+v\,E_{\bf k}^{\rm -})^{n+T_{\rm n}\over2} [x^2-(E_{\bf k}^{\rm +})^2]^{n-T_{\rm n}\over2}[x-{\mu_{\rm Q}\over2}+\epsilon_{\rm s}({\bf k})]^{n+U_{\rm n}\over2}[x-{\mu_{\rm Q}\over2}-\epsilon_{\rm s}({\bf k})]^{n-U_{\rm n}\over2}}\Big|_{x=v\,E_{\bf k}^{\rm -}}\right.\nonumber\\
&&+{\partial^{{n-T_{\rm n}\over2}-1}_x\over\left({n-T_{\rm n}\over2}-1\right)!}{v[x-{\mu_{\rm Q}\over2}+\epsilon_{\rm l}({\bf k})]^{n+T_{\rm n}\over2}[x-{\mu_{\rm Q}\over2}-\epsilon_{\rm l}({\bf k})]^{n-T_{\rm n}\over2}g\left(vx-v{\mu_{\rm Q}+2\mu_{\rm B}\over6},T,L\right)\over 2(x+v\,E_{\bf k}^{\rm +})^{n-T_{\rm n}\over2} [x^2-(E_{\bf k}^{\rm -})^2]^{n+T_{\rm n}\over2}[x-{\mu_{\rm Q}\over2}+\epsilon_{\rm s}({\bf k})]^{n+U_{\rm n}\over2}[x-{\mu_{\rm Q}\over2}-\epsilon_{\rm s}({\bf k})]^{n-U_{\rm n}\over2}}\Big|_{x=v\,E_{\bf k}^{\rm +}}\nonumber\\
&&-{\partial^{{n+U_{\rm n}\over2}-1}_x\over\left({n+U_{\rm n}\over2}-1\right)!}{[x+\epsilon_{\rm l}({\bf k})]^{n+T_{\rm n}\over2}[x-\epsilon_{\rm l}({\bf k})]^{n-T_{\rm n}\over2}g\left(-x+{-\mu_{\rm Q}+\mu_{\rm B}\over3},T,L\right)\over 4 [(x+{\mu_{\rm Q}\over2})^2-(E_{\bf k}^{\rm -})^2]^{n-T_{\rm n}\over2} [(x+{\mu_{\rm Q}\over2})^2-(E_{\bf k}^{\rm +})^2]^{n-T_{\rm n}\over2}[x-\epsilon_{\rm s}({\bf k})]^{n-U_{\rm n}\over2}}\Big|_{x=-\epsilon_{\rm s}({\bf k})}\nonumber\\
&&\left.+{\partial^{{n-U_{\rm n}\over2}-1}_x\over\left({n-U_{\rm n}\over2}-1\right)!}{[x+\epsilon_{\rm l}({\bf k})]^{n+T_{\rm n}\over2}[x-\epsilon_{\rm l}({\bf k})]^{n-T_{\rm n}\over2}g\left(x-{-\mu_{\rm Q}+\mu_{\rm B}\over3},T,L\right)\over 4 [(x+{\mu_{\rm Q}\over2})^2-(E_{\bf k}^{\rm -})^2]^{n-T_{\rm n}\over2} [(x+{\mu_{\rm Q}\over2})^2-(E_{\bf k}^{\rm +})^2]^{n-T_{\rm n}\over2}[x+\epsilon_{\rm s}({\bf k})]^{n+U_{\rm n}\over2}}\Big|_{x=\epsilon_{\rm s}({\bf k})}\right\}.
\eea
Correspondingly, the coefficient $\beta_{\rm 2n}'$ can be separated into vacuum and medium parts, that is, $\beta_{\rm 2n}'\equiv\beta_{\rm 2n}'^{\rm V}+\beta_{\rm 2n}'^{\rm M}$ with
\bea
\beta_{\rm 2n}'^{\rm V}&\equiv&\int{\di^3{\bf k}\over(2\pi)^3}\sum_{t_1,u_2;;t_{\rm n},u_{\rm n}=\pm}\alpha_{t_1,u_1;;t_n,u_n}\xi_{\rm 2n}^{\rm V},\\
\beta_{\rm 2n}'^{\rm M}&\equiv&\int{\di^3{\bf k}\over(2\pi)^3}\sum_{t_1,u_2;;t_{\rm n},u_{\rm n}=\pm}\alpha_{t_1,u_1;;t_n,u_n}\xi_{\rm 2n}^{\rm M}.
\eea
Keep in mind that $\xi_{\rm 2n}^{\rm V/M}$ depends on both $t_{\rm i}$ and $u_{\rm i}$. Usually, $\beta_{\rm 2n}'^{\rm M}$ is convergent as $\lim_{k\rightarrow\infty}k^3\xi_{\rm 2n}^{\rm M}=0$, while $\beta_{\rm 2n}'^{\rm V}$ is divergent and can be regularized by three-momentum cutoff.

\end{widetext}

\end{document}